\documentclass{article}

\usepackage[final]{nips_2016}

\usepackage[utf8]{inputenc} 
\usepackage[T1]{fontenc}    
\usepackage{amsfonts}       
\usepackage{nicefrac}       
\usepackage{microtype}      

\usepackage{siunitx}
\usepackage{hyperref,url,booktabs,microtype, amsfonts,nicefrac,amsmath,bm,algorithm,amsthm,graphicx,tikz,pifont,syntonly, natbib}  
\usepackage{algorithm}
\usepackage[noend]{algpseudocode}
\newcolumntype{L}[1]{>{\raggedright\let\newline\\\arraybackslash\hspace{0pt}}m{#1}}

\title{Efficient and accurate extraction of \emph{in vivo} calcium signals from microendoscopic video data}

\author{
Pengcheng Zhou  \\
Carnegie Mellon University, Columbia University\\
\texttt{pz2230@columbia.edu} \\
\And
Shanna L. Resendez, Jose Rodriguez-Romaguera, Garret D. Stuber \\
University of North Carolina at Chapel Hill \\
\And
Jessica C. Jimenez, Rene Hen\\
Columbia University
\And 
Mazen A. Kheirbek\\
University of California, San Francisco\\
\And 
Shay Q. Neufeld, Bernardo L. Sabatini\\
Harvard University
\AND 
Robert E. Kass\\
Carnegie Mellon University \\
\texttt{kass@stat.cmu.edu}\\
\And 
Liam Paninski\\
Columbia University \\
\texttt{liam@stat.columbia.edu} \\
}

\begin{document}

\maketitle

\begin{abstract}
\emph{In vivo} calcium imaging through microendoscopic lenses enables imaging of previously inaccessible neuronal populations deep within the brains of freely moving animals. However, it is computationally challenging to extract single-neuronal activity from microendoscopic data, because of the very large background fluctuations and high spatial overlaps intrinsic to this recording modality. Here, we describe a new matrix factorization approach to accurately separate the background and then demix and denoise the neuronal signals of interest. We compared the proposed method against widely-used independent components analysis and constrained nonnegative matrix factorization approaches. On both simulated and experimental data, our method substantially improved the quality of extracted cellular signals and detected more well-isolated neural signals, especially in noisy data regimes. These advances can in turn significantly enhance the statistical power of downstream analyses, and ultimately improve scientific conclusions derived from microendoscopic data.
\end{abstract}

\section{Introduction}
Monitoring the activity of large-scale neuronal ensembles during complex behavioral states is fundamental to neuroscience research. Continued advances in optical imaging technology are greatly expanding the size and depth of neuronal populations that can be visualized. Specifically, \emph{in vivo} calcium imaging through microendoscopic lenses and the development of miniaturized microscopes have enabled deep brain imaging of previously inaccessible neuronal populations of freely moving mice \citep{Flusberg2008,Ghosh2011,Ziv2015}. The technique has been widely used to study the neural circuits in cortical, subcortical, and deep brain areas, such as  hippocampus \citep{Cai2016,Rubin2015,Ziv2013}, entorhinal cortex \citep{Kitamura2015,Sun2015}, hypothalamus \citep{Jennings2015}, prefrontal cortex (PFC) \citep{Pinto2015}, premotor cortex \citep{Markowitz2015}, dorsal pons \citep{Cox2016}, basal forebrain \citep{Harrison2016}, striatum \citep{Barbera2016,CarvalhoPoyraz2016,Klaus2017}, amygdala \citep{Yu2017}, and other brain regions.

\begin{figure}[!t]
  \includegraphics[width=1\textwidth]{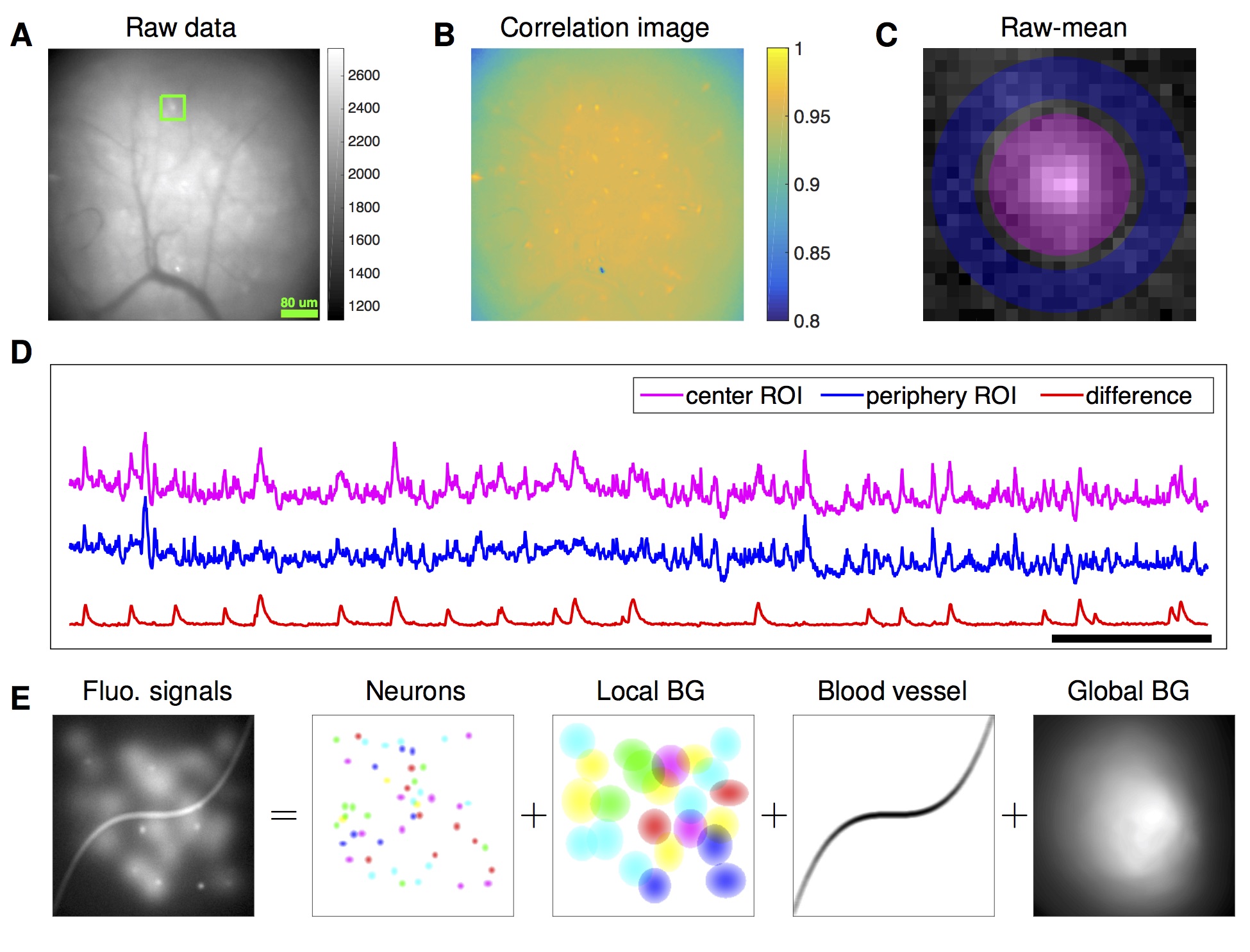}
  \caption{Microendoscopic data contain large background signals with rapid fluctuations due to multiple sources. (\textbf{A}) An example frame of microendoscopic data recorded in dorsal striatum (see Methods and Materials section for experimental details). (\textbf{B}) The local ``correlation image" \citep{Smith2010} computed from the raw video data.  Note that it is difficult to discern neuronal shapes in this image due to the high background spatial correlation level. (\textbf{C}) The mean-subtracted data within the cropped area (green) in (\textbf{A}). Two ROIs were selected and coded with different colors. (\textbf{D}) The mean fluorescence traces of pixels within the two selected ROIs (magenta and blue) shown in (\textbf{C}) and the difference between the two traces. (\textbf{E}) Cartoon illustration of various sources of fluorescence signals in microendoscopic data. ``BG'' abbreviates ``background.''}
  \label{fig:intro}
\end{figure}

Although microendoscopy has potential applications across numerous neuroscience fields \citep{Ziv2015}, methods for extracting cellular signals from this data are currently limited and suboptimal. Most existing methods are specialized for 2-photon or light-sheet microscopy. However, these methods are not suitable for analyzing single-photon microendoscopic data because of its distinct features: specifically, this data typically displays large, blurry background fluctuations due to fluorescence contributions from neurons outside the focal plane.  In Figure \ref{fig:intro} we use a typical microendoscopic dataset to illustrate these effects (see \href{http://www.columbia.edu/~pz2230/videos/example_microendoscopic_data.mp4}{S1 Video} for raw video). Figure \ref{fig:intro}A shows an example frame of the selected data, which contains large signals additional to the neurons visible in the focal plane. These extra fluorescence signals contribute as background that contaminates the single-neuronal signals of interest. In turn, standard methods based on local correlations for visualizing cell outlines \citep{Smith2010} are not effective here, because the correlations in the fluorescence of nearby pixels are dominated by background signals (Figure \ref{fig:intro}B). For some neurons with strong visible signals, we can manually draw  regions-of-interest (ROI) (Figure \ref{fig:intro}C).
Following \citep{Barbera2016,Pinto2015}, we used the mean fluorescence trace of the surrounding pixels (blue, Figure \ref{fig:intro}D) to roughly estimate this background fluctuation; subtracting it from the raw trace in the neuron ROI yields a relatively good estimation of neuron signal (red, Figure \ref{fig:intro}D). Figure \ref{fig:intro}D shows that the background (blue) has much larger variance than the relatively sparse neural signal (red); moreover, the background signal fluctuates on similar timescales as the single-neuronal signal, so we can not simply temporally filter the background away after extraction of the mean signal within the ROI.  This large background signal is likely due to a combination of local fluctuations resulting from out-of-focus fluorescence or neuropil activity, hemodynamics of blood vessels, and global fluctuations shared more broadly across the field of view (photo-bleaching effects, drifts in $z$ of the focal plane, etc.), as illustrated schematically in Figure \ref{fig:intro}E. 

The existing methods for extracting individual neural activity from microendoscopic data can be divided into two classes: semi-manual ROI analysis \citep{Barbera2016,Klaus2017,Pinto2015} and PCA/ICA analysis \citep{Mukamel2009}. Unfortunately, both approaches have well-known flaws \citep{Resendez2016}.  For example, ROI analysis does not effectively demix signals of spatially overlapping neurons, and drawing ROIs is laborious for large population recordings. More importantly, in many cases the background contaminations are not adequately corrected, and thus the extracted signals are not sufficiently clean enough for involved downstream analyses.  As for PCA/ICA analysis, it is a linear demixing method and therefore typically fails when the neural components exhibit strong spatial overlaps \citep{Pnevmatikakis2016} - as is the case in the microendoscopic setting. 

Recently, constrained nonnegative matrix factorization (CNMF) approaches were proposed to simultaneously denoise, deconvolve, and demix calcium imaging data \citep{Pnevmatikakis2016}. However, current implementations of the CNMF approach were optimized for $2$-photon and light-sheet microscopy, where the background has a simpler spatiotemporal structure. When applied to microendoscopic data, CNMF often has poor performance because the background is not modeled sufficiently accurately  \citep{Barbera2016}.  

In this paper, we significantly extend the CNMF framework to obtain a robust approach for extracting single-neuronal signals from microendoscopic data. Specifically, our extended CNMF for microendoscopic data (CNMF-E) approach utilizes a more accurate and flexible spatiotemporal background model that is able to handle the properties of the strong background signal illustrated in Fig.~\ref{fig:intro}, along with new specialized algorithms to initialize and fit the model components. After a brief description of the model and algorithms, we first use simulated data to illustrate the power of the new approach. Next, we compare CNMF-E with PCA/ICA analysis comprehensively on both simulated data and four experimental datasets recorded in different brain areas. The results show that CNMF-E outperforms PCA/ICA in terms of detecting more well-isolated neural signals,  extracting higher signal-to-noise ratio (SNR) cellular signals, and obtaining more robust results in low SNR regimes. Finally, we show that downstream analyses of calcium imaging data can substantially benefit from these improvements.   

\section{Model and model fitting}

\subsection{CNMF for microendoscope data (CNMF-E)}
The recorded video data can be represented by a matrix $Y\in \mathbb{R}_+^{d\times T}$, where $d$ is the number of pixels in the field of view and $T$ is the number of frames observed. In our model each neuron $i$ is characterized by its spatial ``footprint'' vector $\bm{a}_i\in \mathbb{R}_+^d$ characterizing the cell's shape and location, and ``calcium activity'' timeseries $\bm{c}_i\in \mathbb{R}_+^T$, modeling (up to a multiplicative and additive constant) cell $i$'s mean fluorescence signal at each frame.  Here, both $\bm{a}_i$ and $\bm{c}_i$ are constrained to be nonnegative because of their physical interpretations. The background fluctuation is represented by a matrix ${B}\in \mathbb{R}_+^{d\times T}$. If the field of view contains a total number of $K$ neurons, then the observed movie data is modeled as a superposition of all neurons' spatiotemporal activity, plus time-varying background and additive noise:  
\begin{equation}
  Y = \sum_{i=1}^K\bm{a}_i\cdot\bm{c}_i^T +B+E=AC +B+E, \label{eq:Y}
\end{equation}
where  $A=[\bm{a}_1, \ldots, \bm{a}_K]$ and  $C=[\bm{c}_1, \ldots, \bm{c}_K]^T$. The noise term $E\in\mathbb{R}^{d\times T}$ is modeled as Gaussian,  $E(t)\sim \mathcal{N}(\bm{0}, \Sigma)$. $\Sigma$ is a diagonal matrix, indicating that the noise is spatially and temporally uncorrelated.

Estimating the model parameters $A, C$ in model (\ref{eq:Y}) gives us all neurons' spatial footprints and their denoised temporal activity. This can be achieved by minimizing the residual sum of squares (RSS), aka the Frobenius norm of the matrix $Y-(AC+B)$, 
\begin{equation}
  \|Y-(AC+B)\|_F^2,
\end{equation} 
while requiring the model variables $A, C$ and $B$ to follow the desired constraints, discussed below.

\begin{table}[!t]
  \centering
  \begin{tabular}{llll}
    \toprule
    \cmidrule{1-2}
    Name     & Description     &  Domain \\
    \midrule
    $d$ & number of pixels &  $\mathbb{N}_+$  \\
    $T$ & number of frames &  $\mathbb{N}_+$  \\
    $K$ & number of neurons &  $\mathbb{N}$  \\
    $Y$ & motion corrected video data & $\mathbb{R}_+^{d\times T}$     \\
    $A$ & spatial footprints of all neurons  & $\mathbb{R}_+^{d\times K}$   \\
    $C$ & temporal activities of all neurons & $\mathbb{R}_+^{K\times T}$  \\
    $B$ & background activity & $\mathbb{R}_+^{d\times T}$ \\
    $E$ & observation noise & $\mathbb{R}^{d\times T}$  \\
    $W$ & weight matrix to reconstruct $B$ using neighboring pixels& $\mathbb{R}^{d\times d}$  \\
    $\bm{b}_0$ & constraint baseline for all pixels & $\mathbb{R}_+^{d}$  \\
    $\bm{x}_i$ & spatial location of the $i$th pixel & $\mathbb{N}^2$ \\
    $\sigma_i$ & standard deviation of the noise at  pixel $\bm{x}_i$ & $\mathbb{R}_+$  \\
    \bottomrule
  \end{tabular} 
    \caption{Variables used in the CNMF-E model and algorithm. $\mathbb{R}$: real numbers; $\mathbb{R}_+$: positive real numbers; $\mathbb{N}$: natural numbers; $\mathbb{N}_+$: positive integers. }
  \label{table:variables and notations}
\end{table}

\subsubsection{Constraints on neuronal spatial footprints \texorpdfstring{$A$}{Lg} and neural temporal traces \texorpdfstring{$C$}{Lg}}
Each spatial footprint $\bm{a}_i$ should be spatially localized and sparse, since a given neuron will cover only a small fraction of the field of view, and therefore most elements of $\bm{a}_i$ will be zero.  Thus we need to incorporate spatial locality and sparsity constraints on $A$ \citep{Pnevmatikakis2016}. We discuss details further below.

Similarly, the temporal components $\bm{c}_i$ are highly structured, as they represent the cells' fluorescence responses to sparse, nonnegative trains of action potentials.
Following \citep{Vogelstein2010,Pnevmatikakis2016}, we model the calcium dynamics of each neuron $\bm{c}_i$ with a stable autoregressive (AR) process of order $p$,
\begin{equation}
  c_i(t) = \sum_{j=1}^p\gamma_j^{(i)} c_i(t-j)+s_i(t),  \label{eq:arp}
\end{equation}
where $s_i(t)\geq 0$ is the number of spikes that neuron fired at the $t$-th frame.  (Note that there is no further noise input into $c_i(t)$ beyond the spike signal $s_i(t)$.) The AR coefficients $\{\gamma_j^{(i)}\}$ are different for each neuron and they are estimated from the data. In practice, we usually pick $p=2$, thus incorporating both a nonzero rise and decay time of calcium transients in response to a spike; then Eq. (\ref{eq:arp}) can be expressed in matrix form as 
\begin{equation}
  G_i\cdot\bm{c}_i = \bm{s}_i, \text{ with } G_i=\begin{bmatrix}
1 &  0& 0 & \cdots &0 \\ 
-\gamma_1^{(i)}& 1 & 0 & \cdots  & 0\\ 
 -\gamma_2^{(i)}& -\gamma_1^{(i)} &1  &  \cdots& 0\\ 
\vdots  &  \ddots & \ddots &\ddots &\vdots \\ 
 0& \cdots &  -\gamma_2^{(i)}&  -\gamma_1^{(i)}&1 
\end{bmatrix}.
\label{eq:ar2}
\end{equation}
The neural activity $\bm{s}_i$ is nonnegative and typically sparse; to enforce sparsity we can
penalize the $\ell _0$ \citep{Jewell2017} or $\ell _1$ \citep{Pnevmatikakis2016,Vogelstein2010} norm of $\bm{s}_i$, or limit the minimum size of nonzero spike counts \citep{Friedrich2017}. When the rise time constant is small compared to the timebin width (low imaging frame rate), we typically use a simpler AR(1) model (with an instantaneous rise following a spike) \citep{Pnevmatikakis2016}. 

\subsubsection{Constraints on background activity \texorpdfstring{$B$}{Lg}}
Constraints on the background term $B$ in Eq. (\ref{eq:Y}) are essential to the success of CNMF-E, since clearly, if $B$ is completely unconstrained we could just absorb the observed data $Y$ entirely into $B$, which would lead to recovery of no neural activity. At the same time, we need to prevent the residual of the background term (i.e., $B-\hat{B}$, where $\hat{B}$ denotes the estimated spatiotemporal background) from corrupting the estimated neural signals $AC$ in model (\ref{eq:Y}), since subsequently, the extracted neuronal activity would be mixed with background fluctuations, leading to artificially high correlations between nearby cells. This problem is even worse in the microendoscopic context because the background fluctuation usually has significantly larger variance than the isolated cellular signals of interest (Figure \ref{fig:intro}D), and therefore any small errors in the estimation of $B$ can severely corrupt the estimated neural signal $AC$. 

In \citep{Pnevmatikakis2016}, $B$ is modeled as a rank-$1$ nonnegative matrix $B=\bm{b}\cdot \bm{f}^T$, where $\bm{b}\in \mathbb{R}_+^d$ and $\bm{f}\in \mathbb{R}_+^T$. This model mainly captures the global fluctuations within the field of view (FOV). In its application to 2-photon or light-sheet data, this rank-1 model has been shown to be sufficient for relatively small spatial regions; the simple low-rank model does not hold for larger fields of view, and so we can simply divide large FOVs into smaller patches for largely-parallel processing \citep{Friedrich2016,Pnevmatikakis2016}. (See \citep{Pachitariu2016} for an alternative approach.) However, as we will see below, the local rank-1 model fails in many microendoscopic datasets, where multiple large overlapping background sources exist even within modestly-sized FOVs. 

Thus we propose a new model to constrain the background term $B$. We first decompose the background into two terms:
\begin{equation}
  B = B^f + B^c,   \label{eq:B_decompose}
\end{equation}
where $B^f$ represents fluctuating activity and  $B^c=\bm{b}_0\cdot \bm{1}^T$ models constant baselines ($\bm{1}\in\mathbb{R}^T$ denotes a vector of $T$ ones). To model $B^f$, we exploit the fact that background sources (largely due to blurred out-of-focus fluorescence) are empirically much coarser spatially than the average neuron soma size $l$. Thus we model $B^f$ at one pixel as a linear combination of its neighboring pixels' background activities, 
\begin{equation}
  B_{it}^f=\sum_{j\in \Omega_i} w_{ij}\cdot B^f_{jt}, ~ \forall t=1\ldots T, \label{eq:cnmfe_Bi}
\end{equation}
where $\Omega_i = \{j ~ | ~ \text{dist}(\bm{x}_i,\bm{x}_j) \in [l_n, l_n+1)\}$ and $\text{dist}(\bm{x}_i, \bm{x}_j)$ is the Euclidean distance between pixel $i$ and $j$. Thus $\Omega_i$ only selects the neighboring pixels with a distance of $l_n$ from the $i$-th pixel; here $l_n$ is a parameter that we choose to be greater than $l$, e.g., $l_n= 2 l$. 

We can rewrite Eq. (\ref{eq:cnmfe_Bi}) in matrix form: 
\begin{equation}
    B^f = W B^f ,        \label{eq:cnmfe_B}
\end{equation}
where $W_{ij}= 0 $ if $\text{dist}(\bm{x}_i,\bm{x}_j) \notin [l_n, l_n+1)$. 

\subsection{Fitting the CNMF-E model} \label{sec:model_fitting}
Now we can formulate the estimation of all model variables as one optimization problem: 
\begin{align}
\underset{A,C,W, \bm{b}_0}{\min}~& \|Y-AC - \bm{b}_0\cdot \bm{1}^T- B^f\|_F^2 \label{eq:opt_all}\tag{P-All}\\
\text{s.t.}~& A\geq 0, ~A \text{ is sparse and local} \notag\\
 &\bm{c}_i\geq 0,~ \bm{s}_i\geq 0, ~G^{(i)}\bm{c}_i = \bm{s}_i, ~ \bm{s}_i \text{ is sparse}~\forall i=1\ldots K\notag\\
& B^f\cdot \bm{1}= \bm{0}\notag \\
& B^f = W\cdot (Y-AC-\bm{b}_0\cdot \bm{1}^T) \label{eq:Bf}\\
&W_{ij}=0 ~ \text{ if dist}(\bm{x}_i, \bm{x}_j)\notin [l_n, l_n+1). \notag
\end{align}
In this optimization problem, we do not explicitly describe the sparsity constraints of $A$ and $S=[\bm{s}_1, \ldots, \bm{s}_K]^T$ because they can be customized by users under different assumptions (see details in Methods and Materials). In addition, the model variable $B^f$ is not optimized explicitly but can be 
estimated as $W\cdot (Y-AC-\bm{b}_0\cdot \bm{1}^T)$,  and we optimize $W$ instead.  In Eq. (\ref{eq:Bf}), we replace $B^f$ in the right-hand side of Eq. (\ref{eq:cnmfe_B}) with $(Y-AC -\bm{b}_0\cdot \bm{1}^T)$. According to Eq. (\ref{eq:Y}) and (\ref{eq:B_decompose}), this change ignores the noise term $E$. Since elements in $E$ are spatially uncorrelated, $W \cdot E$ contributes as a very small disturbance to our estimated $\hat{B}^f$, which is the left-hand side of Eq. (\ref{eq:Bf}). 

The problem (\ref{eq:opt_all}) optimizes all variables together and is jointly non-convex, but can be divided into three simpler subproblems that we solve iteratively. 

\textbf{Estimating $A$ given $\hat{C}, \hat{B}^f, \hat{\bm{b}}_0$}:
\begin{align}
\underset{A}{\min}~ & \|Y-{A}\cdot \hat{C} - \hat{\bm{b}}_0\cdot \bm{1}^T- \hat{B}^f\|_F^2 \label{eq:opt_A}\tag{P-S}\\
\text{s.t.}~& A\geq 0, ~A \text{ is sparse and local} \notag
\end{align}

\textbf{Estimating $C$ given $\hat{B}^f, \hat{\bm{b}}_0, \hat{A}$}:
\begin{align}
\underset{C}{\min}~ & \|Y-\hat{A}\cdot {C} - \hat{\bm{b}}_0\cdot \bm{1}^T- \hat{B}^f\|_F^2 \label{eq:opt_C}\tag{P-T}\\
\text{s.t.}~ &\bm{c}_i\geq 0, ~\bm{s}_i\geq 0\notag\\
&G^{(i)}\bm{c}_i = \bm{s}_i, ~ \bm{s}_i \text{ is sparse}~ \forall i=1\ldots K\notag
\end{align}

\textbf{Estimating $B^f, {\bm{b}}_0$ given $\hat{A}, \hat{C}$ }
\begin{align}
\underset{W, \bm{b}_0}{\min}~ & \|Y-\hat{A}\cdot \hat{C} - \bm{b}_0\cdot \bm{1}^T- B^f\|_F^2 \label{eq:opt_B}\tag{P-B}\\
\text{s.t.}~ & B^f\cdot \bm{1} = \bm{0}\notag \notag \\
&B^f = W\cdot (Y-\hat{A}\cdot \hat{C}-{\bm{b}}_0\cdot \bm{1}^T).\notag \\
&W_{ij}=0 ~ \text{ if dist}(\bm{x}_i, \bm{x}_j)\notin [l_n, l_n+1)\notag
\end{align}
For each of these subproblems, we are able to use well-established algorithms (e.g., solutions for (\ref{eq:opt_A}) and (\ref{eq:opt_C}) are discussed in \citep{Friedrich2016,Pnevmatikakis2016}) or slight modifications thereof. By iteratively solving these three subproblems, we obtain tractable updates for all model variables in problem (\ref{eq:opt_all}). Furthermore, this strategy gives us the flexibility of further potential interventions (either automatic or semi-manual) in the optimization procedure, e.g., incorporating further prior information on neurons' morphology, or merging/splitting/deleting spatial components and detecting missed neurons from the residuals. These steps can significantly improve the quality of the model fitting; this is an advantage compared with  PCA/ICA, which offers no easy option for incorporation of stronger prior information or manually-guided improvements on the estimates. 

Full details on the algorithms for initializing and then solving these three subproblems are provided in the Methods and Materials section. 

\section{Results} \label{sec:results}
\subsection{CNMF-E can reliably estimate large high-rank background fluctuations}

\begin{figure}[!t]
  \includegraphics[width=1\textwidth]{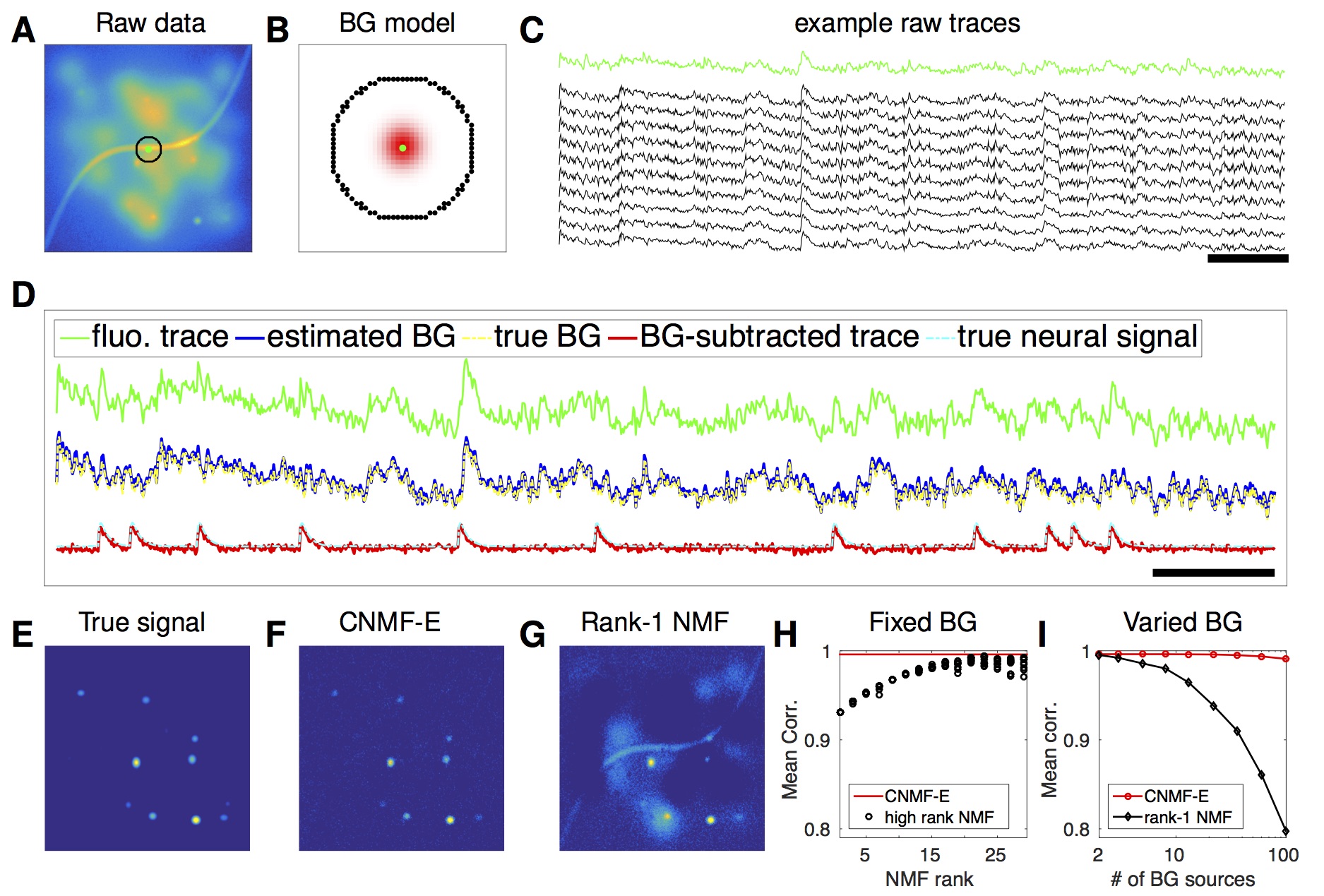}
  \caption{CNMF-E can accurately separate and recover the background fluctuations in simulated data. (\textbf{A}) An example frame of simulated microendoscopic data formed by summing up the fluorescent signals from the multiple sources illustrated in Figure \ref{fig:intro}\textbf{E}. (\textbf{B}) A zoomed-in version of the circle in (\textbf{A}). The green dot indicates the pixel of interest. The surrounding black pixels are its neighbors with a distance of $15$ pixels. The red area approximates the size of a typical neuron in the simulation.  (\textbf{C}) Raw fluorescence traces of the selected pixel and some of its neighbors on the black ring.  Note the high correlation. (\textbf{D}) Fluorescence traces (raw data; true and estimated background; true and initial estimate of neural signal) from the center pixel as selected in (\textbf{B}). Note that the background dominates the raw data in this pixel, but nonetheless we can accurately estimate the background and subtract it away here.  Scalebars: $10$ seconds.  Panels (\textbf{E}-\textbf{G}) show the cellular signals in the same frame as (\textbf{A}). (\textbf{E}) Ground truth neural activity. (\textbf{F}) The residual of the raw frame after subtracting the background estimated with CNMF-E; note the close correspondence with \textbf{E}. (\textbf{G}) Same as (\textbf{F}), but the background is estimated with rank-1 NMF. A video showing (\textbf{E}-\textbf{G}) for all frames can be found at  \href{http://www.columbia.edu/~pz2230/videos/background_comparison.mp4}{S2 Video}. (\textbf{H}) The mean correlation coefficient (over all pixels) between the true background fluctuations and the estimated background fluctuations. The rank of NMF varies and we run randomly-initialized NMF for $10$ times for each rank. The red line is the performance of CNMF-E, which requires no selection of the NMF rank. (\textbf{I}) The performance of CNMF-E and rank-1 NMF in recovering the background fluctuations from the data superimposed with an increasing number of background sources.}
  \label{fig:bg}
\end{figure}

We first use simulated data to illustrate the background model in CNMF-E and compare its performance against the low-rank NMF model used in the basic CNMF approach \citep{Pnevmatikakis2016}. 
We generated the observed fluorescence $Y$ by summing up simulated fluorescent signals of multiple sources as shown in Figure \ref{fig:intro}E plus additive Gaussian white noise (Figure \ref{fig:bg}A). 

An example pixel (green dot, Figure \ref{fig:bg}A,B) was selected to illustrate the background model in CNMF-E  (Eq. (\ref{eq:cnmfe_Bi})), which assumes that each pixel's background activity can be reconstructed using its neighboring pixels' activities. The selected neighbors form a ring and their distances to the center pixel are larger than a typical neuron size  (Figure \ref{fig:bg}B). Figure \ref{fig:bg}C shows that the fluorescence traces of the center pixel and its neighbors are highly correlated due to the shared large background fluctuations. Here for illustrative purposes we fit the background by solving problem (\ref{eq:opt_B}) directly while assuming $\hat{A}\hat{C}=0$. This mistaken assumption should make the background estimation more challenging (due to true neural components getting absorbed into the background), but nonetheless in Figure \ref{fig:bg} we see that the background fluctuation was well recovered (Figure \ref{fig:bg}D). Subtracting this estimated background from the observed fluorescence in the center yields a good visualization of the cellular signal (Figure \ref{fig:bg}D). Thus this example shows that we can reconstruct a complicated background trace while leaving the neural signal uncontaminated. 

For the example frame in Figure \ref{fig:bg}A, the true cellular signals are sparse and weak (Figure \ref{fig:bg}E). When we subtract the estimated background using CNMF-E from the raw data, we obtain a good recovery of the true signal (Figure \ref{fig:bg}D,F). For comparison, we also estimate the background activity by applying a rank-$1$  NMF model as used in basic CNMF; the resulting background-subtracted image is still severely contaminated by the background (Figure \ref{fig:bg}G). This is easy to understand: the spatiotemporal background signal in microendoscopic data typically has a rank higher than one, due to the various signal sources indicated in Figure \ref{fig:intro}E), and therefore a rank-$1$ NMF background model is insufficient. 

A naive approach would be to simply increase the rank of the NMF background model.  Figure \ref{fig:bg}H demonstrates that this approach is ineffective: higher-rank NMF does yield generally better reconstruction performance, but with high variability and low reliability (due to randomness in the initial conditions of NMF).  Eventually as the NMF rank increases many single-neuronal signals of interest are swallowed up in the estimated background signal (data not shown).  In contrast, CNMF-E recovers the background signal more accurately than any of the high-rank NMF models.

In real data analysis settings, the rank of NMF is an unknown  and the selection of its value is a nontrivial problem. We simulated data sets with different numbers of local background sources and use a single parameter setting to run CNMF-E for reconstructing the background over multiple such simulations. Figure \ref{fig:bg}I shows that the performance of CNMF-E does not degrade quickly as we have more background sources, in contrast to  rank-$1$ NMF. Therefore CNMF-E can recover the background accurately across a diverse range of background sources, as desired.  

\subsection{CNMF-E accurately initializes single-neuronal spatial and temporal components} \label{sec:init_results}

\begin{figure}[!t]
  \includegraphics[width=1\textwidth]{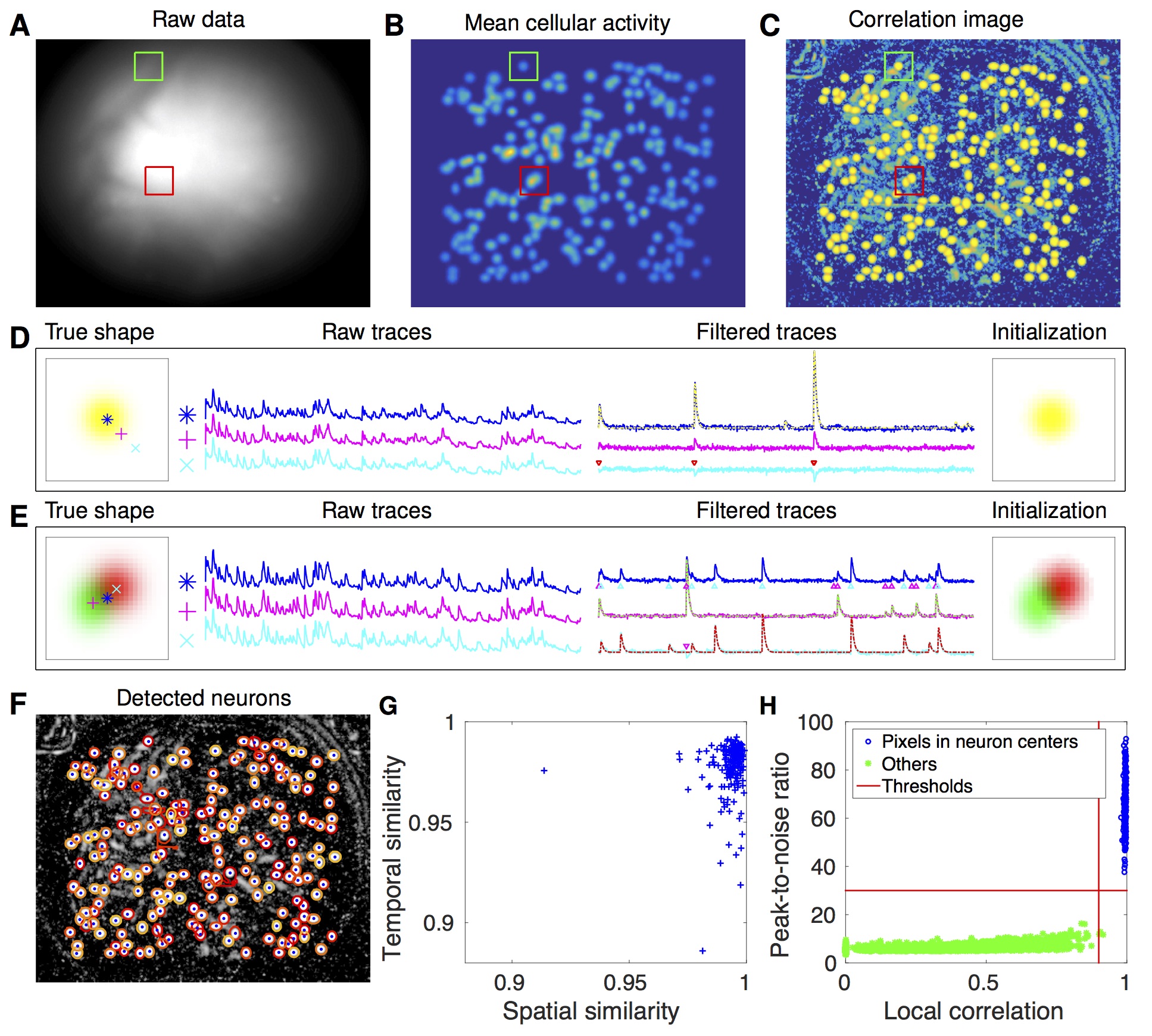}
  \caption{CNMF-E accurately initializes individual neurons' spatial and temporal components in simulated data. (\textbf{A}) An example frame of the simulated data. Green and red squares will correspond to panels (\textbf{D}) and (\textbf{E}) below, respectively. (\textbf{B}) The temporal mean of the cellular activity in the simulation. (\textbf{C}) The correlation image computed using the spatially filtered data. (\textbf{D}) An example of initializing an isolated neuron. Three selected pixels correspond to the center, the periphery, and the outside of a neuron. The raw traces and the filtered traces are shown as well. The yellow dashed line is the true neural signal of the selected neuron. Triangle markers highlight the spike times from the neuron. (\textbf{E}) Same as (\textbf{D}), but two neurons are spatially overlapping in this example.  Note that in both cases neural activity is clearly visible in the filtered traces, and the initial estimates of the spatial footprints are already quite accurate (dashed lines are ground truth). (\textbf{F}) The contours of all initialized neurons on top of the correlation image as shown in (\textbf{D}). Contour colors represent the rank of neurons' SNR (SNR decreases from red to yellow). The blue dots are centers of the true neurons.  (\textbf{G}) The spatial and the temporal cosine similarities between each simulated neuron and its counterpart in the initialized neurons. (\textbf{H}) The local correlation and the peak-to-noise ratio for pixels located in the central area of each neuron (blue) and other areas (green). The red lines are the thresholding boundaries for screening seed pixels in our initialization step. A video showing the whole initialization step can be found at  \href{http://www.columbia.edu/~pz2230/videos/sim_initialization.mp4}{S3 Video}.}
  \label{fig:fig_init}
\end{figure}

Next we used simulated data to validate our proposed initialization procedure (Figure \ref{fig:fig_init}A). 
In this example we simulated $200$ neurons with strong spatial overlaps (Figure \ref{fig:fig_init}B). One of the first steps in our initialization procedure is to apply a Gaussian spatial filter to the images to reduce the (spatially coarser) background and boost the power of neuron-sized objects in the images. In Figure \ref{fig:fig_init}C, we see that the local correlation image \citep{Smith2010} computed on the spatially filtered data provides a good initial visualization of neuron locations; compare to Figure \ref{fig:intro}B, where the correlation image computed on the raw data was highly corrupted by background signals. 

We choose two example ROIs to illustrate how CNMF-E removes the background contamination and demixes nearby neural signals for accurate initialization of neurons' shapes and activity. In the first example, we choose a well-isolated neuron (green box, Figure \ref{fig:fig_init}A+B). We select three pixels located in the center, the periphery, and the outside of the neuron and show the corresponding fluorescence traces in both the raw data and the spatially filtered data (Figure \ref{fig:fig_init}D). The raw traces are noisy and highly correlated, but the filtered traces show relatively clean neural signals. This is because spatial filtering reduces the shared background activity and the remaining neural signals dominate the filtered data. 
Similarly, Figure \ref{fig:fig_init}E is an example showing how CNMF-E demixes two overlapping neurons. The filtered traces in the centers of the two neurons still preserve their own temporal activity.

After initializing the neurons' traces using the spatially filtered data, we initialize our estimate of their spatial footprints; in this case the initial values already match the simulated ground truth with high fidelity (Figure \ref{fig:fig_init}D+E).
In this simulated data, CNMF-E successfully identified all $200$ neurons and initialized their spatial and temporal components (Figure \ref{fig:fig_init}F). We then evaluate the quality of initialization using all neurons' spatial and temporal similarities with their counterparts in the ground truth data. Figure \ref{fig:fig_init}G shows that all initialized neurons have high similarities with the truth, indicating a good recovery and demixing of all neuron sources.

Thresholds on the minimum local correlation and the minimum peak-to-noise ratio (PNR) for detecting seed pixels are necessary for defining the initial spatial components. 
 To quantify the sensitivity of choosing these two thresholds, we plot the local correlations and the PNRs of all pixels  chosen as the local maxima within an area of $\frac{l}{4}\times \frac{l}{4}$, where $l$ is the diameter of a typical neuron,  in the correlation image or the PNR image (Figure \ref{fig:fig_init}H). Pixels are classified into two classes according to their  locations relative to the closest neurons: neurons' central areas  and outside areas (see Methods and Materials for full details).
 It is clear that the two classes are  linearly well separated and the thresholds can be chosen within a broad range of values (Figure \ref{fig:fig_init}H), indicating that the algorithm is robust with respect to these threshold parameters. 

\subsection{CNMF-E recovers the true neural activity and is robust to noise contaminations on simulated data}
\begin{figure}[!t]
  \centering
  \includegraphics[width=1\textwidth]{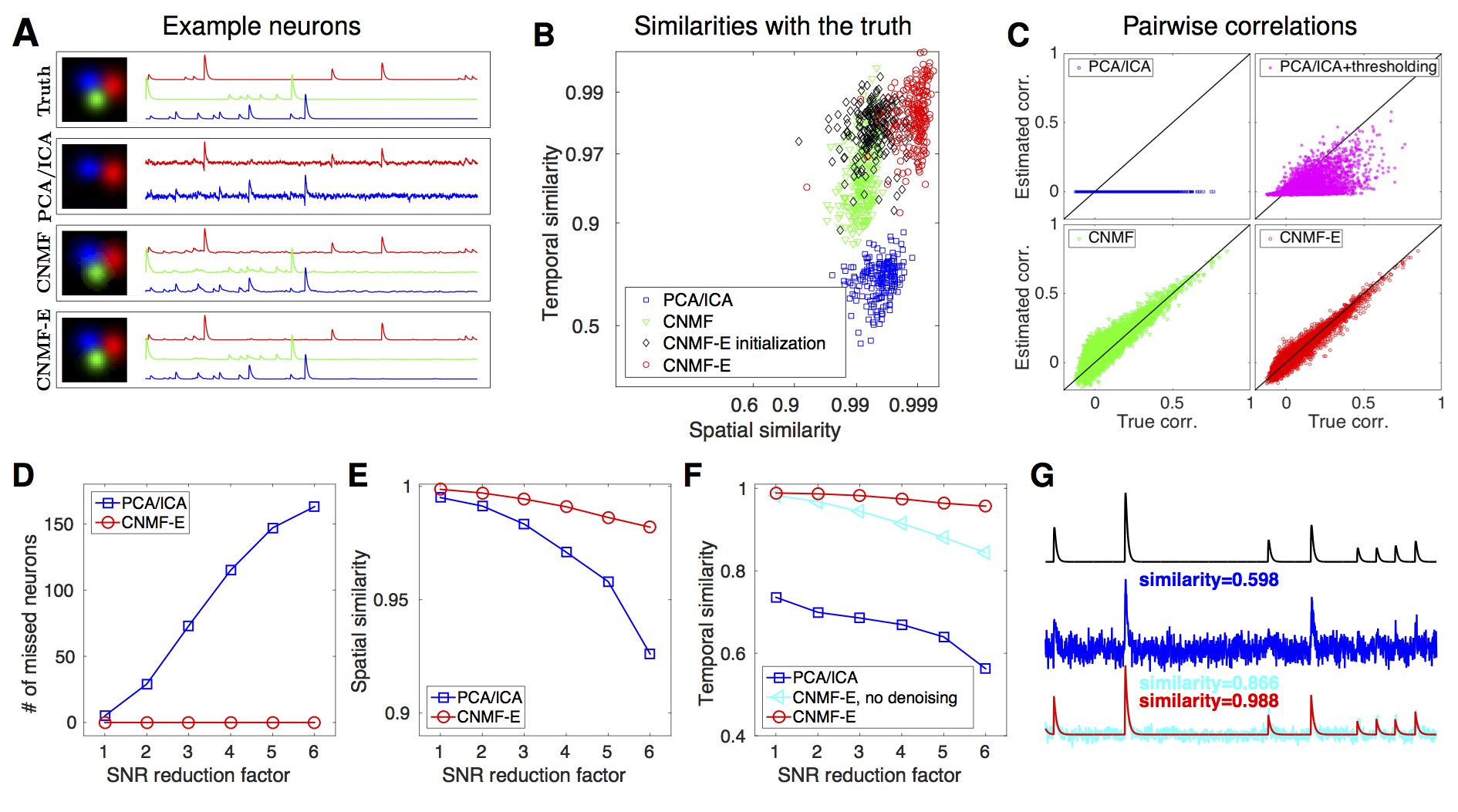}
  \caption{CNMF-E outperforms PCA/ICA analysis in extracting individual neurons' activity from simulated data and is robust to low SNR. (\textbf{A}) The results of PCA/ICA, CNMF, and CNMF-E in recovering the spatial footprints and temporal traces of three example neurons. The trace colors match the neuron colors shown in the left. (\textbf{B}) The spatial and the temporal cosine similarities between the ground truth and the neurons detected using different methods. (\textbf{C}) The pairwise correlations between the calcium activity traces extracted using different methods.  (\textbf{D}-\textbf{F}) The performances of PCA/ICA and CNMF-E under different noise levels: the number of missed neurons (\textbf{D}), and the spatial (\textbf{E})  and temporal (\textbf{F}) cosine similarities between the extracted components and the ground truth. (\textbf{G}) The calcium traces of one example neuron: the ground truth (black), the PCA/ICA trace (blue), the CNMF-E trace (red) and the CNMF-E trace without being denoised (cyan). The similarity values shown in the figure are computed as the cosine similarity between each trace and the ground truth (black). Two videos showing the demixing results of the simulated data can be found in  \href{http://www.columbia.edu/~pz2230/videos/sim_snr1_demixing.mp4}{S4 Video}  (SNR reduction factor=1)  and  \href{http://www.columbia.edu/~pz2230/videos/sim_snr6_demixing.mp4}{S5 Video} (SNR reduction factor=6).}
  \label{fig:sim_compare}
\end{figure}

Using the same simulated dataset as in the previous section, we further refine the neuron shapes ($A$) and the temporal traces ($C$) by iteratively fitting the CNMF-E model.  We compare the final results with  PCA/ICA analysis \citep{Mukamel2009} and the original CNMF method \citep{Pnevmatikakis2016}. 

After choosing the thresholds for seed pixels (Figure \ref{fig:fig_init}H), we run CNMF-E  in full automatic mode, without any manual interventions. Two open-source MATLAB packages, CellSort \footnote{https://github.com/mukamel-lab/CellSort} and ca\_source\_extraction \footnote{https://github.com/epnev/ca\_source\_extraction}, were used  to perform PCA/ICA \citep{Mukamel2009} and basic CNMF \citep{Pnevmatikakis2016}, respectively. Since the initialization algorithm in the CNMF fails due to the large contaminations from the background fluctuations in this setting (recall Figure \ref{fig:bg}), we use the ground truth as its initialization. As for the rank of the background model in CNMF, we tried all integer values between $1$ and $16$ and set it as $9$ because it has the best performance in matching the ground truth. We emphasize that including the CNMF approach in this comparison is not fair for the other two approaches, because it uses the ground truth heavily, while PCA/ICA and CNMF-E are blind to the ground truth. The purpose here is to show the limitations of basic CNMF in modeling the background activity in microendoscopic data.

We first pick three closeby neurons from the ground truth (Figure \ref{fig:sim_compare}A, top) and see how well these neurons' activities are recovered. PCA/ICA fails to identify one neuron, and for the other two identified neurons, it recovers temporal traces that are sufficiently noisy that small calcium transients are submerged in the noise. As for CNMF, the neuron shapes remain more or less at the initial condition (i.e., the ground truth spatial footprints), but clear contaminations in the temporal traces are visible. This is because the pure NMF model in CNMF does not model the true background well and the residuals in the background are mistakenly captured by neural components. In contrast, on this example, CNMF-E recovers the true neural shapes and neural activity with high accuracy. 

We also compare the number of detected neurons and how well these neurons are detected. We detected $195$ out of $200$ neurons using PCA/ICA, while CNMF-E detected all $200$ neurons. We also quantitatively evaluated the performance of source extraction by showing the spatial and temporal cosine similarities between detected neurons and ground truth (Figure \ref{fig:sim_compare}B). As a comparison, the neurons detected using PCA/ICA have much lower similarities with the ground truth (Figure \ref{fig:sim_compare}B). We also note that CNMF results are much worse than CNMF-E, despite the fact that CNMF is initialized at the ground truth parameter values. Compared with the results in the initialization step, running the whole pipeline of CNMF-E leads to improvements in both spatial and temporal similarities.

In many downstream analyses of calcium imaging data, pairwise correlations provide an important metric to study coordinated network activity \citep{Barbera2016,Dombeck2009,Klaus2017,Warp2012}. Since PCA/ICA seeks statistically independent components, which forces the temporal traces to have near-zero correlation, the correlation structure is badly corrupted in the raw PCA/ICA outputs (Figure \ref{fig:sim_compare}C). We observed that a large proportion of the independence comes from the noisy baselines in the extracted traces (data not shown), so we postprocessed the PCA/ICA output by thresholding at the $3$ standard deviation level. This recovers some nonzero correlations, but the true correlation structure is not recovered accurately (Figure \ref{fig:sim_compare}C). By contrast, the CNMF-E results matched the ground truth very well due to accurate extraction of individual neurons' temporal activity (Figure \ref{fig:sim_compare}C). As for CNMF, the estimated correlations are slightly elevated relative to the true correlations. This is due to the shared (highly correlated) background fluctuations that corrupt the recovered activity of nearby neurons. 

Finally, we compare the performance of the different methods under different SNR regimes.  Because of the above inferior results we skip comparisons to the basic CNMF here. Based on the same simulation parameters as above, we vary the noise level $\Sigma$ by multiplying it with a SNR reduction factor. Figure \ref{fig:sim_compare}D shows that CNMF-E detects all neurons over a wide SNR range, while PCA/ICA fails to detect the majority of neurons when the SNR drops to sufficiently low levels. Moreover, the detected neurons in CNMF-E preserve high  spatial and  temporal similarities with the ground truth (Figure \ref{fig:sim_compare}E-F). This high accuracy of extracting neurons' temporal activity benefits from the modeling of the calcium dynamics, which leads to significantly denoised neural activity.  If we skip the temporal denoising step in the algorithm, CNMF-E is less robust to noise, but still outperforms PCA/ICA significantly (Figure \ref{fig:sim_compare}F). When SNR is low, the improvements yielded by CNMF-E can be crucial for detecting weak neuron events, as shown in Figure \ref{fig:sim_compare}G.

\subsection{Application to dorsal striatum data}

\begin{figure}[!t]
  \centering
  \includegraphics[width=1\textwidth]{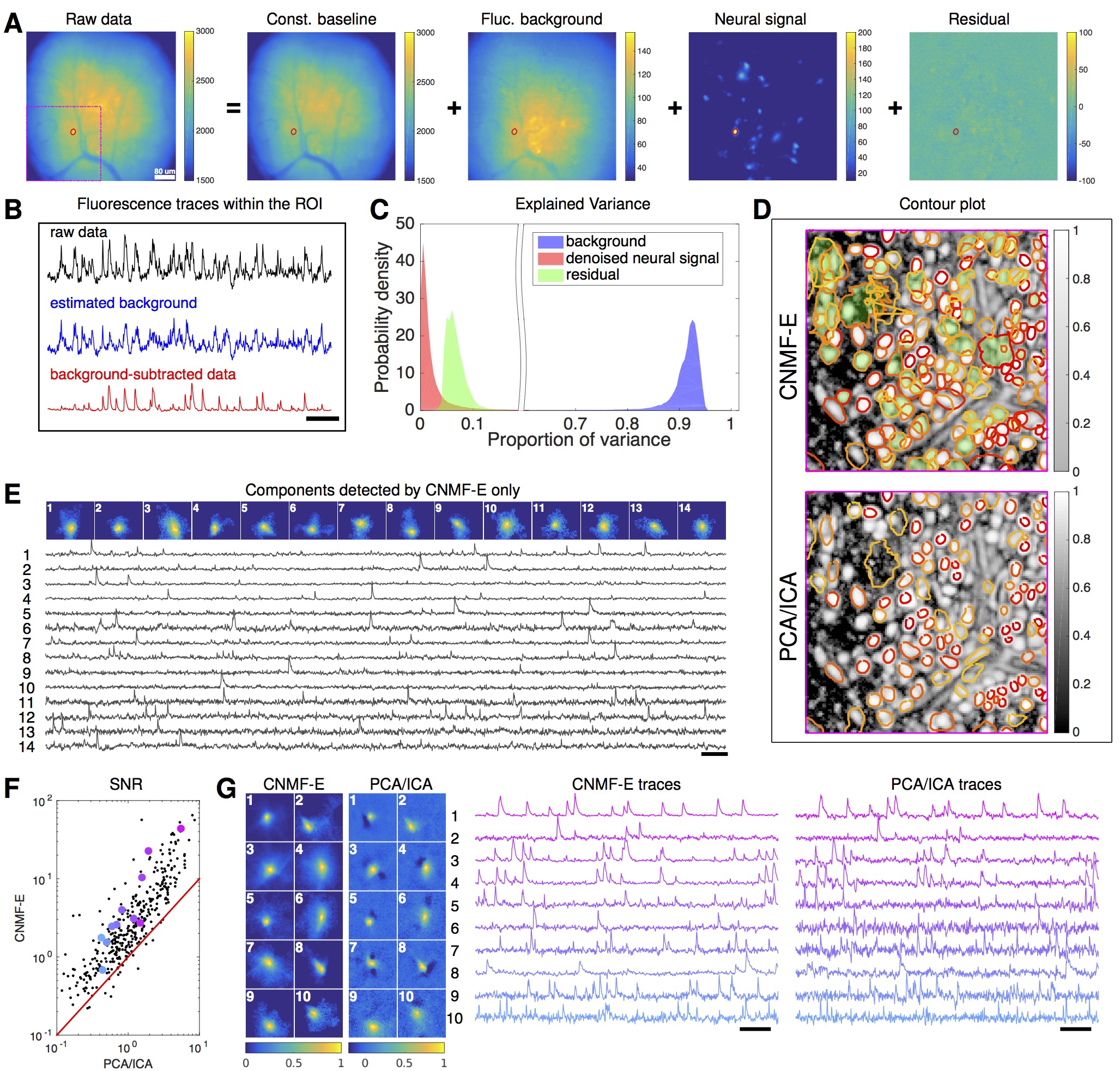}
  \caption{Neurons expressing GCaMP6f recorded \emph{in vivo} in mouse dorsal striatum area. (\textbf{A}) An example frame of the raw data and its four components decomposed by CNMF-E. (\textbf{B}) The mean fluorescence traces of the raw data (black), the estimated background activity (blue), and the background-subtracted data (red) within the segmented area (red) in (\textbf{A}). The variance of the black trace is about 2x the variance of the blue trace and 4x the variance of the red trace.  (\textbf{C}) The distributions of the variance explained by different components over all pixels; note that estimated background signals dominate the total variance of the signal. (\textbf{D}) The contour plot of all neurons detected by CNMF-E and PCA/ICA superimposed on the correlation image. Green areas represent the components that are only detected by CNMF-E. The components are sorted in decreasing order based on their SNRs (from red to yellow). (\textbf{E})  The spatial and temporal components of $14$ example neurons that are only detected by CNMF-E. These neurons all correspond to green areas in (\textbf{D}). (\textbf{F}) The signal-to-noise ratios (SNRs) of all neurons detected by both methods. Colors match the example traces shown in (\textbf{G}), which shows the spatial and temporal components of $10$ example neurons detected by both methods. Scalebar: $10$ seconds. See \href{http://www.columbia.edu/~pz2230/videos/striatum_demixing.mp4}{S6 Video} for the demixing results. } 
  \label{fig:dorsal_striatum}
\end{figure}

We now turn to the analysis of large-scale microendoscopic datasets recorded from freely behaving mice. We run both CNMF-E and PCA/ICA for all datasets and compare their performances in detail. 

We begin by analyzing \emph{in vivo} calcium imaging data of neurons expressing GCaMP6f in the mouse dorsal striatum.  (Full experimental details and algorithm parameter settings for this and the following datasets appear in the Methods and Materials section.)
CNMF-E extracted 550 putative neural components from this dataset; PCA/ICA extracted 384 components (starting from 700 initial components, and then manually removing independent components whose spatial filter appeared to consist of random pixels or whose temporal traces had no prominent calcium events).
Figure \ref{fig:dorsal_striatum}A shows how CNMF-E decomposes an example frame into four components: the constant baselines that are invariant over time, the fluctuating background, the denoised neural signals, and the residuals. We highlight an example neuron by drawing its ROI to demonstrate the power of  CNMF-E in isolating  fluorescence signals of neurons from the background fluctuations. For the selected neuron, we plot the mean fluorescence trace of the raw data and the estimated background (Figure \ref{fig:dorsal_striatum}B). These two traces are very similar, indicating that the background fluctuation dominates the raw data. By subtracting this estimated background component from the raw data, we acquire a clean trace that represents the neural signal. 

To quantify the background effects further, we compute the contribution of each signal component in explaining the variance in the raw data. For each pixel, we compute the variance of the raw data first and then compute the variance of the background-subtracted data. Then the reduced variance is divided by the variance of the raw data, giving the proportion of variance explained by the background. Figure \ref{fig:dorsal_striatum}C (blue) shows the distribution of the background-explained variance over all pixels. The background accounts for around $90\%$ of the variance on average. We further remove the denoised neural signals and compute the variance reduction; Figure \ref{fig:dorsal_striatum}C shows that neural signals account for less than $10\%$ of the raw signal variance. This analysis is consistent with our observations that background dominates the fluorescence signal and  extracting high-quality neural signals requires careful background signal removal.  

The contours of the spatial footprints inferred by the two approaches (PCA/ICA and CNMF-E) are depicted in Figure \ref{fig:dorsal_striatum}D, superimposed on the correlation image of the filtered raw data. The indicated area was cropped from Figure \ref{fig:dorsal_striatum}A (left). 
In this case, all neurons inferred by PCA/ICA were inferred by CNMF-E as well. However, many components were only detected by CNMF-E (shown as the green areas in Figure \ref{fig:dorsal_striatum}D). In these plots, we rank the inferred components according to their SNRs; the color indicates the relative rank (decaying from red to yellow). We see that the components missed by PCA/ICA have low SNRs (green shaded areas with yellow contours). 

Figure \ref{fig:dorsal_striatum}E shows the spatial and temporal components of $14$ example neurons detected only by CNMF-E. Here (and in the following figures), for illustrative purposes, we show the calcium traces before the temporal denoising step. 
For neurons that are inferred by both methods, CNMF-E shows significant improvements in the SNR of the extracted cellular signals (Figure \ref{fig:dorsal_striatum}F), even before the temporal denoising step is applied.  In panel G we randomly select $10$ examples and examine their spatial and temporal components. Compared with the CNMF-E results, PCA/ICA components have much smaller size, often with negative dips surrounding the neuron (remember that ICA avoids spatial overlaps in order to reduce nearby neurons' statistical dependences, leading to some loss of signal strength; see \citep{Pnevmatikakis2016} for further discussion). The activity traces extracted  by CNMF-E are visually cleaner than the PCA/ICA traces; this is important for reliable event detection, particularly in low SNR examples. See \citep{Klaus2017} for additional examples of  CNMF-E applied to striatal data.

\subsection{Application to data in prefrontal cortex}

\begin{figure}[!t]
  \centering
  \includegraphics[width=1\textwidth]{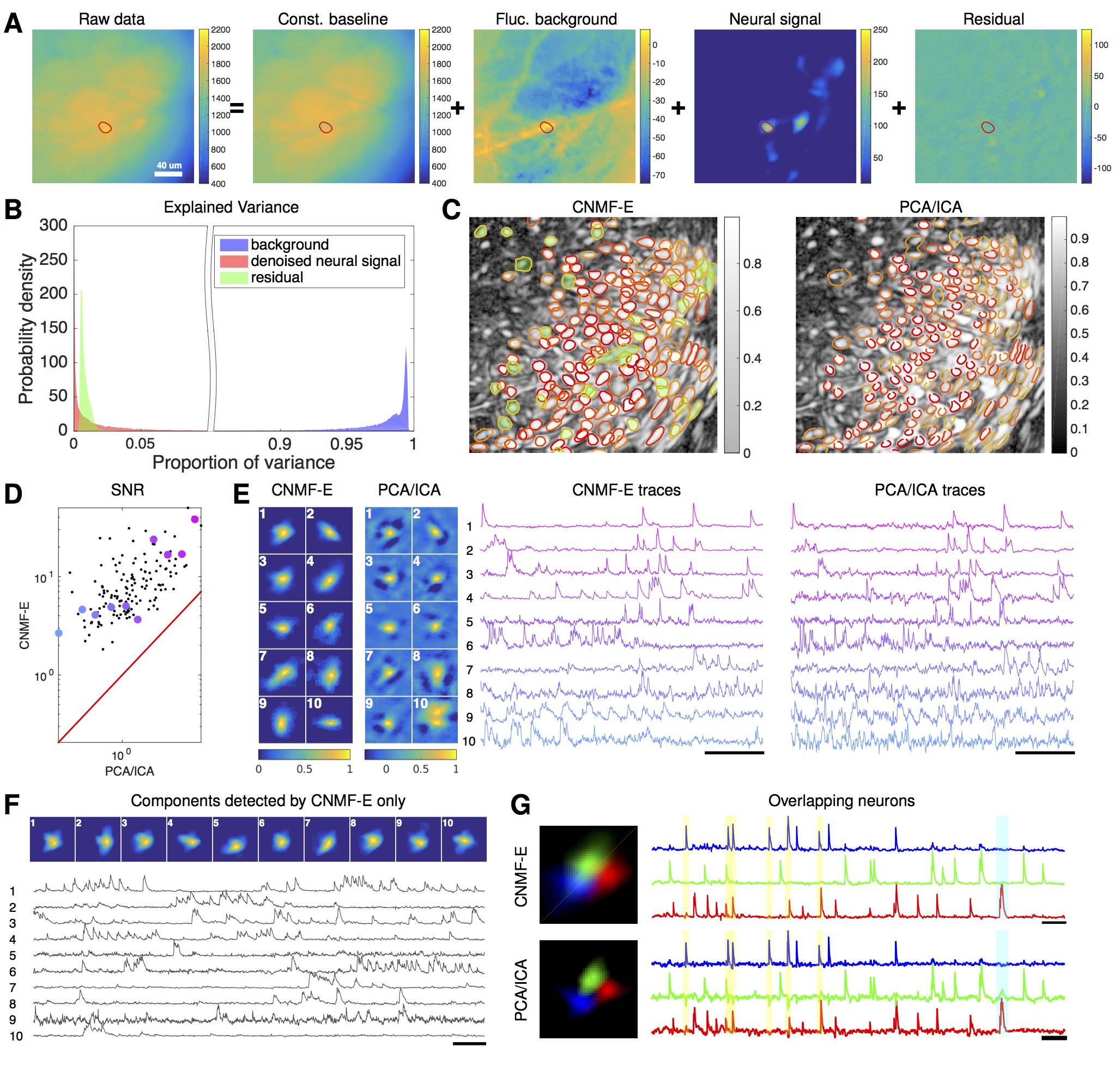}
  \caption{Neurons expressing GCaMP6s  recorded \emph{in vivo} in mouse prefrontal cortex. (\textbf{A}-\textbf{F}) follow similar conventions as in the corresponding panels of Figure \ref{fig:dorsal_striatum}. (\textbf{G}) Three example neurons that are close to each other and detected by both methods. Yellow shaded areas highlight the negative `spikes' correlated with nearby activity, and the cyan shaded area highlights one crosstalk between nearby neurons. Scalebar: $20$ seconds.  See \href{http://www.columbia.edu/~pz2230/videos/pfc_demixing.mp4}{S7 Video} for the demixing results and \href{http://www.columbia.edu/~pz2230/videos/pfc_overlapping.mp4}{S8 Video} for the comparision of CNMF-E and PCA/ICA in the zoomed-in area of (\textbf{G}).}
  \label{fig:pfc}
\end{figure}

We repeat a similar analysis on GCaMP6s data recorded from prefrontal cortex (PFC, Figure \ref{fig:pfc}), to quantify the performance of the algorithm in a different brain area with a different calcium indicator. Again we find that CNMF-E successfully extracts neural signals from a strong fluctuating background (Figure \ref{fig:pfc}A), which contributes a large proportion of the variance in the raw data (Figure \ref{fig:pfc}B). Similarly as with the striatum data, PCA/ICA analysis missed many components that have very weak signals ($35$ missed components here). For the matched neurons, CNMF-E shows strong improvements in the SNRs of the extracted traces (Figure \ref{fig:pfc}D). Consistent with our observation in striatum (Figure \ref{fig:dorsal_striatum}G), the spatial footprints of PCA/ICA components are shrunk to promote statistical independence between neurons, while the neurons inferred by CNMF-E have visually reasonable morphologies (Figure \ref{fig:dorsal_striatum}E). Some neurons inferred by PCA/ICA fail to appropriately split two nearby neurons (Figure \ref{fig:pfc}E, cell $10$). As for calcium traces with high SNRs (Figure \ref{fig:pfc}E, cell $1-6$), CNMF-E traces have smaller noise values, which is important for detecting small calcium transients (Figure \ref{fig:pfc}E, cell $4$).  For traces with low SNRs (Figure \ref{fig:pfc}, cell $7-10$), it is challenging to detect any calcium events from the PCA/ICA traces due to the large noise variance; CNMF-E is able to visually recover many of these weaker signals.  For those cells missed by PCA/ICA, their traces extracted by CNMF-E have reasonable morphologies and visible calcium events (Figure \ref{fig:pfc}F). 

The demixing performance of PCA/ICA analysis can be relatively weak because it is inherently a linear demixing method \citep{Pnevmatikakis2016}. Since CNMF-E uses a more suitable nonlinear matrix factorization method, it has a better capability of demixing spatially overlapping neurons. As an example, Figure \ref{fig:pfc}G shows three closeby neurons identified by both CNMF-E and PCA/ICA analysis. PCA/ICA forces its obtained filters to be spatially separated to reduce their dependence (thus reducing the effective signal strength), while CNMF-E allows inferred spatial components to have large overlaps (Figure \ref{fig:pfc}G, left), retaining the full signal power. In the traces extracted by PCA/ICA, the component labeled in green contains many negative ``spikes," which are highly correlated with the spiking activity of the blue neuron (Figure \ref{fig:pfc}G, yellow).
In addition, the green PCA/ICA neuron has significant crosstalk  with the red neuron due to the failure of signal demixing (Figure \ref{fig:pfc}G, cyan); the CNMF-E traces shows no comparable negative ``spikes" or crosstalk.  See also \href{http://www.columbia.edu/~pz2230/videos/pfc_overlapping.mp4}{S8 Video} for further details.

\subsection{Application to ventral hippocampus neurons}

\begin{figure}[!t]
  \centering
  \includegraphics[width=1\textwidth]{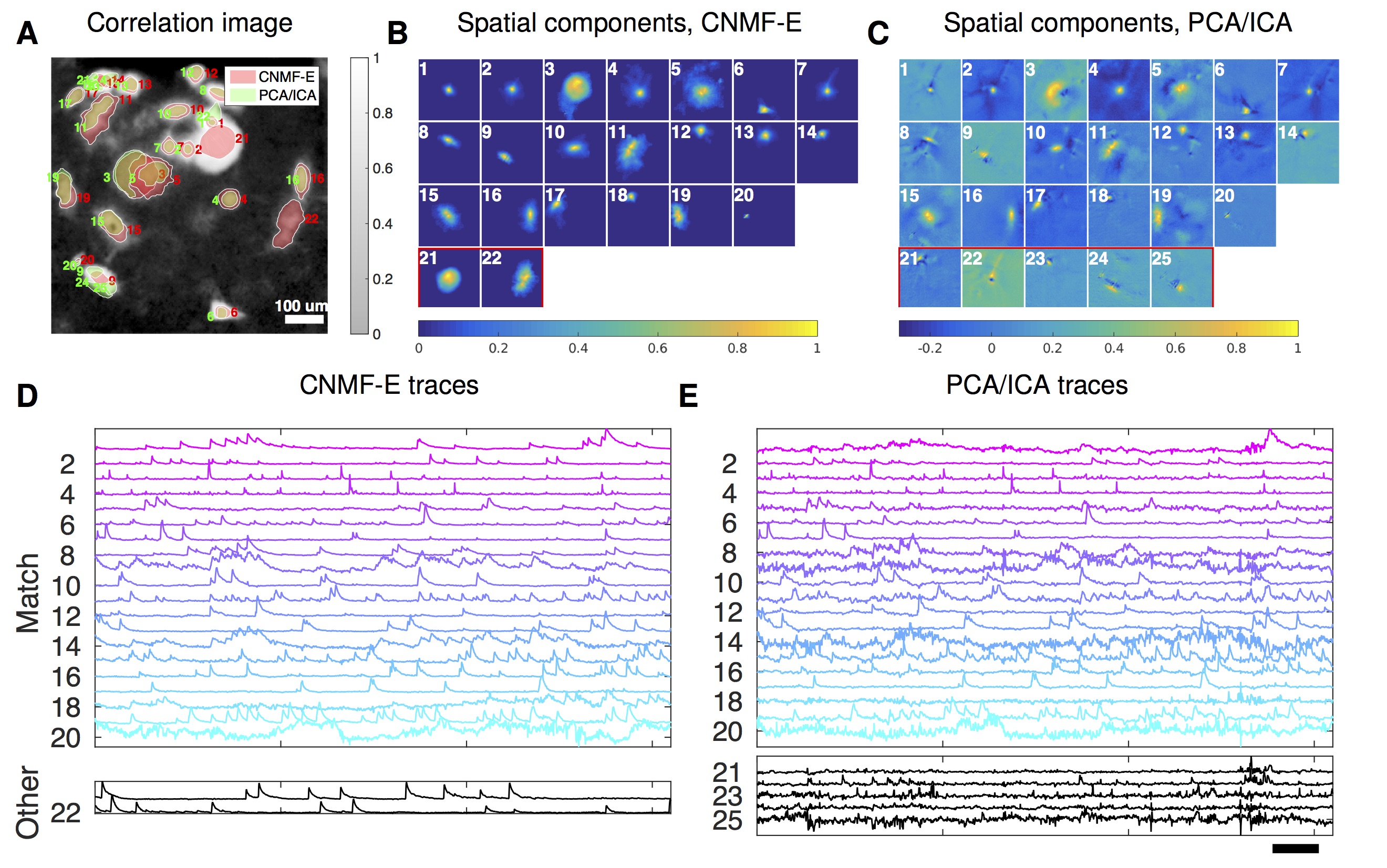}
  \caption{Neurons expressing GCaMP6f recorded \emph{in vivo} in mouse ventral hippocampus. (\textbf{A}) Contours of all neurons detected by CNMF-E (red) and PCA/ICA method (green). The grayscale image is the local correlation image of the background-subtracted video data, with background estimated using CNMF-E. (\textbf{B}) Spatial components of all neurons detected by CNMF-E. The neurons in the first three rows are also detected by PCA/ICA, while the neurons in the last row are only detected by CNMF-E. (\textbf{C}) Spatial components of all neurons detected by PCA/ICA; similar to (\textbf{B}), the neurons in the first three rows are also detected by CNMF-E and the neurons in the last row are only detected by PCA/ICA method. (\textbf{D}) Temporal traces of all detected components in (\textbf{B}). `Match' indicates neurons in top three rows in panel (\textbf{B}); `Other' indicates neurons in the fourth row. (\textbf{E}) Temporal traces of all components in (\textbf{C}). Scalebars: $20$ seconds.  See \href{http://www.columbia.edu/~pz2230/videos/sparse_demixing.mp4}{S9 Video} for demixing results.
}
  \label{fig:sparse}\end{figure}
 
In the previous two examples, we analyzed data with densely packed neurons, in which the neuron sizes are all similar. In the next example, we apply CNMF-E to a dataset with much sparser and more heterogeneous neural signals. The data used here were recorded from amygdala-projecting neurons expressing GCaMP6f in ventral hippocampus. In this dataset, some neurons that are slightly above or below the focal plane were visible with prominent signals, though their spatial shapes are larger than neurons in the focal plane. 

This example is somewhat more challenging due to the large diversity of neuron sizes.  It is possible to set multiple parameters to detect neurons of different sizes (or to e.g. differentially detect somas versus smaller segments of axons or dendrites passing through the focal plane), but for illustrative purposes here we use a single neural size parameter to initialize all of the components. This in turn splits some large neurons into multiple components. Following this crude initialization step, we ran three iterations of updating the model variables $A, C$, and $B$, together with some manual merge/delete interventions (see Methods and Materials  below), leading to improved source extraction results (see   \href{http://www.columbia.edu/~pz2230/videos/intervention_results.mp4}{S10 Video} for details on the manual merge and delete interventions performed here).  In this example, we detected $22$ CNMF-E components and $25$ PCA/ICA components. The contours of these inferred neurons are shown in Figure \ref{fig:sparse}A. In total we have $20$ components detected by both methods (shown in the first three rows of Figure \ref{fig:sparse}B+C); each method detected extra components that are not detected by the other (the last rows of Figure \ref{fig:sparse}B+C). Once again, the PCA/ICA filters contain many negative pixels in an effort to reduce spatial overlaps; see components $3$ and $5$ in Figure \ref{fig:sparse}A-C, for example. All traces of the inferred neurons are shown in Figure \ref{fig:sparse}D+E. We can see that the CNMF-E traces have much lower noise level and cleaner neural signals in both high and low SNR settings. Conversely, the calcium traces of the 5 extra neurons identified by PCA/ICA show noisy signals that are unlikely to be neural responses. 

\subsection{Application to footshock responses in the bed nucleus of the stria terminalis (BNST)}

\begin{figure}[!t]
  \centering
  \includegraphics[width=1\textwidth]{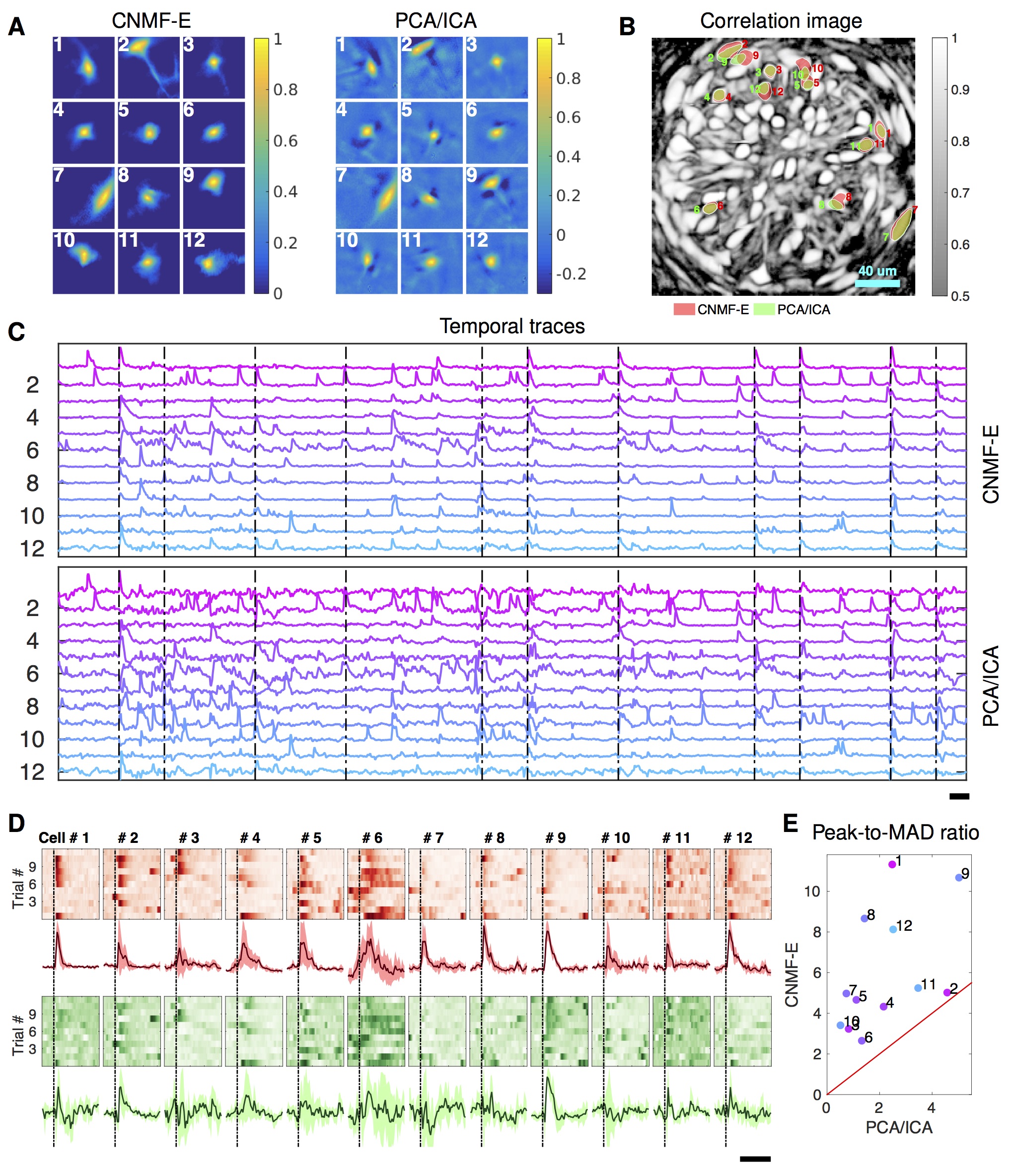}
  \caption{Neurons extracted by CNMF-E show more reproducible responses to footshock stimuli, with larger signal sizes relative to the across-trial variability, compared to PCA/ICA. (\textbf{A}-\textbf{C}) Spatial components (\textbf{A}), spatial locations (\textbf{B}) and temporal components (\textbf{C}) of $12$ example neurons detected by both CNMF-E and PCA/ICA.  (\textbf{D}) Calcium responses of all example neurons  to footshock stimuli. Colormaps show trial-by-trial responses of each neuron, extracted by CNMF-E (top, red) and PCA/ICA (bottom, green), aligned to the footshock time.  The solid lines are medians of neural responses over $11$ trials and the shaded areas correpond to median $\pm1$ median absolute deviation (MAD). Dashed lines indicate the shock timings. (\textbf{E}) Scatter plot of peak-to-MAD ratios for all response curves in (\textbf{D}). For each neuron, Peak is corrected by subtracting the mean activity within $4$ seconds prior to stimulus onset and MAD is computed as the mean MAD values over all timebins shown in (\textbf{D}). The red line shows $y=x$. Scalebars: $10$ seconds. See \href{http://www.columbia.edu/~pz2230/videos/footshock_demixing.mp4}{S11 Video} for demixing results.
}
  \label{fig:footshock}
\end{figure}

Identifying neurons and extracting their temporal activity is typically just the first step in the analysis of calcium imaging data; downstream analyses rely heavily on the quality of this initial source extraction.  We showed above that, compared to PCA/ICA, CNMF-E is better at extracting activity dynamics, especially in regimes where neuronal activities are correlated (c.f. Figure \ref{fig:sim_compare}C). Using \emph{in vivo} electrophysiological recordings, we previously showed that neurons in the bed nucleus of the stria terminalis (BNST) show strong responses to unpredictable footshock stimuli \citep{Jennings2013}.  We therefore measured calcium dynamics in CaMKII-expressing neurons that were transfected with the calcium indicator GCaMP6s in the BNST and analyzed the synchronous activity of multiple neurons in response to unpredictable footshock stimuli.  We chose $12$ example neurons that were detected by both CNMF-E and PCA/ICA methods and show their spatial and temporal components in Figure \ref{fig:footshock}A-C. The activity around the onset of the repeated stimuli are aligned and shown as pseudo-colored images in panel D. The median responses of CNMF-E neurons display prominent responses to the footshock stimuli compared with the resting state before stimuli onset.  In comparison, the activity dynamics extracted by PCA/ICA have relatively low SNR, making it more challenging to reliably extract footshock responses. Panel E summarizes the results of panel D; we see that CNMF-E outputs significantly more easily detectable responses than does PCA/ICA.  This is an example in which downstream analyses of calcium imaging data can significantly benefit from the improvements in the accuracy of source extraction offered by CNMF-E. 

\section{Conclusion}
Microendoscopic calcium imaging offers unique advantages and has quickly become a critical method for recording large neural populations during unrestrained behavior.  However, previous methods fail to adequately remove background contaminations when demixing single neuron activity from the raw data. Since strong background signals are largely inescapable in the context of one-photon imaging, insufficient removal of the background could yield problematic conclusions in downstream analysis. This has presented a severe and well-known bottleneck in the field.  We have delivered a solution for this critical problem, building on the constrained nonnegative matrix factorization framework introduced in \citet{Pnevmatikakis2016} but significantly extending it in order to more accurately and robustly remove these contaminating background components. 

The proposed CNMF-E algorithm can be used in either automatic or semi-automatic mode, and leads to significant improvements in the accuracy of source extraction compared with previous methods. In addition, CNMF-E requires very few parameters to be specified, and these parameters are easily interpretable and can be selected within a broad range. We demonstrated the power of CNMF-E using  data from a wide diversity of brain areas (subcortical, cortical, and deep brain areas), SNR regimes, calcium indicators, neuron sizes and densities, and hardware setups. Among all these examples (and many others not shown here), CNMF-E performs well and improves significantly on the standard PCA/ICA approach. Further applications of the CNMF-E approach appear in \citep{Cameron2016,Donahue2017,Jimenez2016,Jimenez2017,Klaus2017, Lin2017, Murugan2016,Murugan2017,Rodriguez-Romaguera2017,Tombaz2016,Ung2017,Yu2017,Mackevicius2017,Madangopal2017}.


We have released our MATLAB implementation of CNMF-E as open-source software (\url{https://github.com/zhoupc/CNMF_E}). We welcome additions or suggestions for modifications of the code, and hope that the large and growing microendoscopic imaging community finds CNMF-E to be a helpful tool in furthering neuroscience research.

\section{Methods and Materials}

\subsection{Algorithms for solving problem (\ref{eq:opt_A})} 
\label{supp:update_A}
We let $\tilde{Y}=Y - \hat{\bm{b}}_0\cdot \bm{1}^T- \hat{B}^f$ and rewrite problem (\ref{eq:opt_A}) as 
\begin{align}
\underset{A}{\min}~ & \|\tilde{Y}-{A}\cdot \hat{C}\|_F^2 \label{eq:opt_A_short}\tag{P-S'}\\
\text{s.t.}~ & A\geq 0, ~ \text{A is local and sparse}.\notag 
\end{align}
Two algorithms were used to solve problem (\ref{eq:opt_A_short}) given different constraints on the sparsity of $A$. Here, we briefly describe the formulation of these two algorithms; details  can be found in the referenced papers. 

\subsubsection{HALS}
HALS stands for hierarchical alternating least squares \citep{Cichocki2007}. It is a standard algorithm for nonnegative matrix factorization. \citeauthor{Friedrich2017} modified the fastHALS algorithm \citep{Cichocki2009} to estimate the nonnegative spatial components $A, \bm{b}$ and the nonnegative temporal activity $C, \bm{f}$ in CNMF model $Y=A\cdot C+\bm{b}\bm{f}^T+E$  by including sparsity and localization constraints \citep{Friedrich2016}. When we remove the sparsity constraint from problem (\ref{eq:opt_A_short}), the new problem is exactly the subproblem of the modified HALS in \citep{Friedrich2016}, 
\begin{align}
\underset{A}{\min}~ & \|\tilde{Y}-{A}\cdot \hat{C}\|_F^2 \label{eq:opt_A1}\tag{P-S1}\\
\text{s.t.}~ & A\geq 0, ~ A(i,k)=0 ~ \forall ~ \bm{x}_i\notin P_k\notag 
\end{align}
where $P_k$ denotes the the spatial patch constraining the nonzero pixels of the $k$-th neurons. The spatial patches can be determined using the previous estimation of $A$. 

\subsubsection{LARS}
In the original CNMF paper, \citet{Pnevmatikakis2016} update the sparse matrix $\hat{A}$ by minimizing its $\ell _1$ norm while constraining the residuals at each pixel to be bounded by the noise variance, 
\begin{align}
\underset{A}{\min}~ & \|A\|_1 \label{eq:opt_A_lars}\tag{P-S2}\\
\text{s.t.}~ & A\geq 0, \notag\\
&\|\tilde{Y}(i,:)-A(i,:)\cdot \hat{C}\|\leq \sigma_i\sqrt{T}, ~ \forall i=1\ldots d.\notag 
\end{align}
This new optimization problem is equivalent to (\ref{eq:opt_A_short}) when we add $\sum_{i=1}^d\lambda_i\cdot \|A(i,:)\|_1$ to its objective function $\|\tilde{Y}-{A}\cdot \hat{C}\|_F^2$ for sparseness penalization, where $\lambda_i\geq 0$.  Problem (\ref{eq:opt_A_lars}) is large, but we can update each row of $A$ separately. The nonnegative LARS algorithm is used to solve \ref{eq:opt_A_lars} for each pixel \citep{Pnevmatikakis2016,Efron2004}. 

\subsection{Algorithms for solving problem (\ref{eq:opt_C})} \label{supp:udpate_C}
Similarly as with problem (\ref{eq:opt_A}), we rewrite problem (\ref{eq:opt_C}) as 
\begin{align}
\underset{C}{\min}~ & \|\tilde{Y}-\hat{A}\cdot C\|_F^2 \label{eq:opt_C_simple}\tag{P-T'}\\
\text{s.t.}~ &\bm{c}_i\geq 0, \bm{s}_i\geq 0, ~ G^{(i)}\cdot\bm{c}_i = \bm{s}_i,  ~ \bm{s}_i \text{ is sparse} ~ \forall i=1\ldots K. \notag
\end{align}
Problem (\ref{eq:opt_C}) is convex and a global minimum exists. However, it is expensive to solve due to the large number of constraints. We follow the block coordinate-descent approach used in \citep{Pnevmatikakis2016}. For each neuron, we construct a raw trace $\bm{y}_{i}$ that minimizes the residual of the spatiotemporal data matrix while fixing other neurons' spatiotemporal activity, 
\begin{equation}
  \hat{\bm{y}}_i = \hat{\bm{c}}_i + \frac{\hat{\bm{a}}_i^T\cdot (\tilde{Y}-\hat{A}\cdot\hat{C})}{\hat{\bm{a}}_i^T\hat{\bm{a}}_i}. \label{eq:y_raw}
\end{equation}
Then various deconvolution algorithms can be applied to compute the denoised trace $\hat{\bm{c}}_i$ and deconvolved signal $\hat{\bm{s}}_i$ from $\hat{\bm{y}}_i$ \citep{Friedrich2017,Jewell2017,Pnevmatikakis2013,Park2013, Vogelstein2009,Vogelstein2010}. These algorithms mainly differ in their constraints on the sparsity of $\bm{s}_i$; see the referenced papers for full details. In CNMF-E, we mainly use constrained FOOPSI \citep{Pnevmatikakis2016} or thresholded OASIS \citep{Friedrich2017}.


\subsection{Estimating background by solving problem (\ref{eq:opt_B})} \label{sec:solve_pb}
Next we discuss our algorithm for estimating the spatiotemporal background signal by solving problem (\ref{eq:opt_B}) as a linear regression problem given $\hat{A}$ and $\hat{C}$. Since  $B^f\cdot \bm{1}=\bm{0}$, we can easily estimate the constant baselines for each pixel as 
\begin{equation}
   \hat{\bm{b}}_0 = \frac{1}{T}(Y-\hat{A}\cdot\hat{C})\cdot \bm{1}.
\end{equation}
Next we replace the $\bm{b}_0$ in (\ref{eq:opt_B}) with this estimate and rewrite (\ref{eq:opt_B}) as 
\begin{align}
\underset{W}{\min}~ & \|{X}- W\cdot {X}\|_F^2, \label{opt:W}\tag{P-W}\\
\text{s.t.} ~ & W_{ij}=0 ~ \text{ if } \text{dist}(\bm{x}_i, \bm{x}_j) \notin [l_n, l_n+1) \notag,  
\end{align}
where ${X}=Y-\hat{A}\cdot \hat{C}-\hat{\bm{b}}_0 \bm{1}^T$. Given the optimized $\hat{W}$, our estimation of the fluctuating background is $\hat{B}^f = \hat{W}\tilde{X}$. The new optimization problem (\ref{opt:W}) can be readily parallelized into $d$ linear regression problems  for each pixel separately. By estimating all row columns of $W_{i,:}$, we are able to obtain the whole background signal as 
\begin{equation}
  \hat{B} = \hat{W}X + \hat{\bm{b}}_0\bm{1}^T. 
\end{equation}

In some cases, $X$ might include large residuals from the inaccurate estimation of the neurons' spatiotemporal activity $AC$, e.g., missing neurons in the estimation. These residuals act as outliers and distort the estimation of $\hat{B}^f$ and $\bm{b}_0$. To overcome this problem, we use robust least squares regression (RLSR) via hard thresholding to avoid contaminations from the outliers \citep{bhatia2015robust}. Before solving the problem (\ref{opt:W}), we preprocess $X$ by letting
\begin{equation}
  {X}_{it}=\left\{\begin{matrix}
B_{it}^-& ~ \text{if}~ X_{it}\geq B_{it}^-+\zeta\cdot \sigma_i\\ 
X_{it}& ~ \text{else}
\end{matrix}\right.. 
\end{equation}
where $B^-$ is the current estimation of the fluctuating background. $\sigma_i$ is the standard deviation of the noise at $\bm{x}_i$ and its value can be estimated using the power spectral density (PSD) method \citep{Pnevmatikakis2016}. As for the first iteration of the model fitting, we set each $B_{it}^-=\frac{1}{|\Omega_i|}\sum_{j\in\Omega_i}\tilde{X}_{jt}$ as the mean of the $\tilde{X}_{jt}$ for all $j\in \Omega_i$. The thresholding coefficient $\zeta$ can be specified by users, though we have found a fixed default works well across the datasets used here. This preprocessing removes most calcium transients by replacing those frames with the previously estimated background only. As a result, it increases the robustness to inaccurate estimation of $AC$, and in turn leads to a better extraction of $AC$ in the following iterations. 

\subsection{Initialization of model variables} 

Since problem (\ref{eq:opt_all}) is not jointly convex in all of its variables, a good initialization of model variables is crucial for fast convergence and accurate extraction of all neurons' spatiotemporal activity. Previous methods assume the background component is relatively weak, allowing us to initialize $\hat{A}$ and $\hat{C}$ while ignoring the background or simply initializing it with a constant baseline over time. However, the noisy background in microendoscopic data fluctuates more strongly than the neural signals (c.f. Figure \ref{fig:dorsal_striatum}C and Figure \ref{fig:pfc}B), which makes previous methods  less valid for the initialization of CNMF-E. 

\begin{figure}[!t]
  \includegraphics[width=1\textwidth]{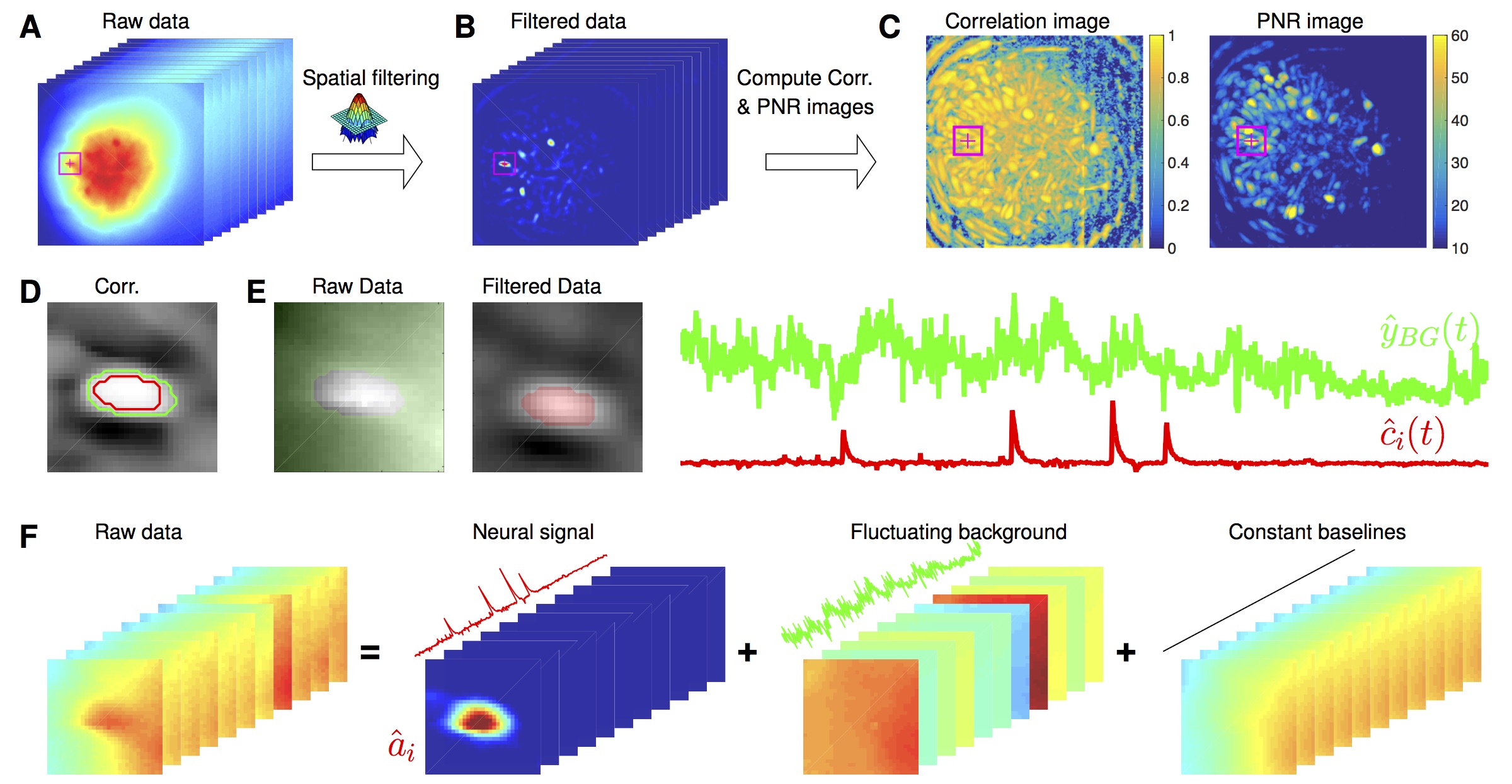}
  \caption{Illustration of the initialization procedure. (\textbf{A}) Raw video data and the kernel for filtering the video data. (\textbf{B}) The spatially high-pass filtered data. (\textbf{C}) The local correlation image and the peak-to-noise ratio (PNR) image calculated from the filtered data in (\textbf{B}). (\textbf{D}) The temporal correlation coefficients between the filtered traces (\textbf{B}) of the selected seed pixel (the red cross) and all other pixels in the cropped area as shown in (\textbf{A}-\textbf{C}). The red and green contour correspond to correlation coefficients equal to $0.7$ and $0.3$ respectively. (\textbf{E}) The estimated background fluctuation $y_{BG}(t)$ (green) and the initialized temporal trace $\hat{c}_i(t)$ of the neuron (red). $y_{BG}(t)$ is computed as the median of the raw fluorescence traces of all pixels (green area) outside of the green contour shown in (\textbf{D}) and $\hat{c}_i(t)$ is computed as the mean of the filtered fluorescence traces of all pixels inside the red contour. (\textbf{F}) The decomposition of the raw video data within the cropped area. Each component is a rank-$1$ matrix and the related temporal traces are estimated in (\textbf{E}). The spatial components are estimated by regressing the raw video data against these three traces. See  \href{http://www.columbia.edu/~pz2230/videos/sim_initialization.mp4}{S3 Video} for an illustration of the initialization procedure.}
  \label{fig:initialization}
\end{figure}

Here we design a new algorithm to initialize $\hat{A}$ and $\hat{C}$ without estimating $\hat{B}$. The whole procedure is illustrated in Figure \ref{fig:initialization} and described in Algorithm \ref{algo:init}. The key aim of our algorithm is to exploit the relative spatial smoothness in the background compared to the single neuronal signals visible in the focal plane. Thus we can use spatial filtering  to  reduce the background in order to estimate single neurons' temporal activity, and then initialize each neuron's spatial footprint given these temporal traces. Once we have $\hat{A}$ and $\hat{C}$, it is straightforward to initialize the constant baseline $\bm{b}_0$ and the fluctuating background $B^f$ by solving problem (\ref{eq:opt_B}). 

\begin{algorithm}[t!]\small 
\caption{Initialize model variables $A$ and $C$ given the raw data}\label{algo:init}
\begin{algorithmic}[1]
\Require data $Y\in \mathbb{R}^{d\times T}$, neuron size $l$, the minimum local correlation $L_{min}$ and the minimum PNR  $P_{min}$ for selecting seed pixels
\State $h(\bm{x}) = \exp\{-\frac{\|\bm{x}\|^2}{2(l/4)^2}\}$; \Comment{Gaussian kernel approximating a typical neuron}
\State $\tilde{h}(\bm{x}) \leftarrow h(\bm{x}) - \bar{h}(\bm{x})$ ; \Comment{kernel for spatial filtering}
\State $Z \leftarrow \text{conv}(Y, h(\bm{x}))$; \Comment{spatially filter the raw data}
\State $L(\bm{x})\leftarrow $ local cross-correlation  image of the filtered data $Z$
\State $P(\bm{x})\leftarrow $  PNR image of the filtered data $Z$
\State $k\leftarrow 0$ \Comment{neuron number}
\While{True}
	\If{$L(\bm{x})\leq L_{min}$ or $P(\bm{x})\leq P_{min}$ for all $\bm{x}$}
	\State break;
  \Else
  \State $k\leftarrow k+1$
	\State $\bm{x}^* \leftarrow \text{argmax}_{\bm{x}}(L(\bm{x})\cdot P(\bm{x}))$; \Comment{select a seed pixel}
  \State $\Omega_k \leftarrow \{\bm{x}|\bm{x} \text{ is in the square box of length }( 2 l+1) \text{ surrounding pixel } \bm{x}^*\}$ \Comment{crop a small box near $\bm{x}^*$}
	\State $\text{corr}(\bm{x}, \bm{x}^*)\leftarrow \text{corr}(z(\bm{x},t), z(\bm{x}^*,t))$ for all $\bm{x}\in\Omega_k$ 
  \State $y_{BG}(t) \leftarrow \frac{\sum_{\text{corr}(\bm{x},\bm{x}^*)\leq 0.3} y(\bm{x},t)}{\sum_{\text{corr}(\bm{x},\bm{x}^*)\leq 0.3}1}$ for all $\bm{x}\in\Omega_k$ \Comment{estimate the background signal using the raw data}
  \State $\hat{{c}}_k(t) \leftarrow \frac{\sum_{\text{corr}(\bm{x},\bm{x}^*)\geq 0.7} z(\bm{x},t)}{\sum_{\text{corr}(\bm{x},\bm{x}^*)\geq 0.7}1}$ for all $\bm{x}\in\Omega_k$ \Comment{estimate neural signal using the filtered data}
  \State $\hat{\bm{a}}_k, \hat{\bm{b}}_f, \hat{\bm{b}}_0 \leftarrow \text{argmin}_{\bm{a}_k, \bm{b}_f, \bm{b}_0} \|Y_{\Omega_k} -(\bm{a}_k\cdot \hat{\bm{c}}_k^T+\bm{b}_f\cdot \bm{y}_{BG}^T+\bm{b}_0\cdot \bm{1}^T))\|^2_2$   
  \State $\hat{\bm{a}}_k = \max(0, \hat{\bm{a}}_k)$ \Comment{the spatial component of $k$-th neuron}
  \State $Y\leftarrow Y - \hat{\bm{a}}_k\cdot \hat{\bm{c}}_k^T$
  \State update $L(\bm{x})$ and $P(\bm{x})$ locally given the new  $Y$
    \EndIf

\EndWhile
\State $A\leftarrow [\bm{a}_1, \bm{a}_2, \hdots, \bm{a}_k]$
\State $C\leftarrow [\bm{c}_1, \bm{c}_2, \hdots, \bm{c}_k]^T$
\State \textbf{return} $A, C$
 \end{algorithmic} 
\end{algorithm}

\subsubsection{Spatially filtering the data}
We first filter the raw video data with a customized image kernel (Figure \ref{fig:initialization}A). The kernel is generated from a Gaussian filter 
\begin{equation}
  h(\bm{x}) = \exp \left( -\frac{\|\bm{x}\|^2}{2(l/4)^2} \right). \label{eq:gauss_kernel}
\end{equation}
Here we use $h(\bm{x})$ to approximate a cell body; the factor of $1/4$ in the Gaussian width is chosen to match a Gaussian shape to a cell of width $l$. Instead of using $h(\bm{x})$ as the filtering kernel directly, we subtract its spatial mean (computed over a region of width equal to  $l$)  and filter the raw data with $\tilde{h}(\bm{x}) = {h}(\bm{x})-\bar{h}(\bm{x}) $. The filtered data is denoted as $Z\in \mathbb{R}^{d\times T}$ (Figure \ref{fig:initialization}B). This spatial filtering step helps accomplish two goals: (1) reducing the background $B$, so that $Z$ is dominated by neural signals (albeit somewhat spatially distorted) in the focal plane (see Figure \ref{fig:initialization}B as an example); (2) performing a template matching to detect cell bodies similar to the Gaussian kernel. Consequently, $Z$ has large values near the center of each cell body.  (However, note that we can not simply e.g. apply CNMF to $Z$, because the spatial components in a factorization of the matrix $Z$ will typically no longer be nonnegative, and therefore NMF-based approaches can not be applied directly.)  More importantly, the calcium traces near the neuron center in the filtered data  preserve the calcium activity of the corresponding neurons because the filtering step results in a weighted average of cellular signals surrounding each pixel (Figure \ref{fig:initialization}B). Thus the fluorescent traces in pixels close to neuron centers in $Z$ can be used for initializing the neurons' temporal activity directly. These pixels are defined as seed pixels. We next propose a quantitative method to rank all potential seed pixels. 

\subsubsection{Ranking seed pixels}
A seed pixel $\bm{x}$ should have two main features: first, $Z(\bm{x})$, which is the filtered trace at pixel $\bm{x}$,  should have high peak-to-noise ratio (PNR) because it encodes the calcium concentration $\bm{c}_i$ of one neuron; second, a seed pixel should have high temporal correlations with  its neighboring pixels (e.g., $4$ nearest neighbors) because they share the same $\bm{c}_i$. We computed two metrics for each of these two features: 
\begin{equation}
  P(\bm{x}) = \frac{\max_t(Z(\bm{x},t))}{\sigma(\bm{x})},~  L(\bm{x})=\frac{1}{4}\sum_{\text{dist}(\bm{x},\bm{x}')=1} \text{corr}\Big(Z(\bm{x}),Z(\bm{x}')\Big). \label{eq:pnr_corr}
\end{equation}
Recall that $\sigma(\bm{x})$ is the standard deviation of the noise at pixel $\bm{x}$; the function $\textbf{corr}()$ refers to Pearson correlation here. In our implementation, we usually threshold  $Z(\bm{x})$ by $3\sigma(\bm{x})$ before computing $L(\bm{x})$ to reduce the influence of the background residuals, noise, and spikes from nearby neurons. 

Most pixels can be ignored directly when selecting seed pixels because their local correlations or PNR values are too small. To  avoid unnecessary searches of the pixels, we set thresholds for both $P(\bm{x})$  and $L(\bm{x})$, and only pick pixels larger than the thresholds $P_{\min}$ and $L_{\min}$. It is empirically useful to combine both metrics for screening seed pixels. For example, high PNR values could result from  large noise, but these pixels usually have small $L(\bm{x})$ because the noise is not shared with neighboring pixels. On the other hand, insufficient removal of background during the spatial filtering leads to high $L(\bm{x})$, but the corresponding $P(\bm{x})$ are usually small because most background fluctuations have been removed. So we create another matrix $R(\bm{x}) = P(\bm{x})\cdot L(\bm{x})$ that computes the pixelwise product of $P(\bm{x})$ and $L(\bm{x}$). We rank all $R(\bm{x})$ in a descending order and choose the pixel $\bm{x}^*$ with the largest $R(\bm{x})$ for initialization.

\subsubsection{Greedy initialization}
Our initialization method greedily initializes neurons one by one. Every time we initialize a neuron, we will remove its initialized spatiotemporal activity from the raw video data and initialize the next neuron from the residual. For the same neuron, there are several seed pixels that could be used to initialize it. But once the neuron has been initialized from any of these seed pixels (and the spatiotemporal residual matrix has been updated by peeling away the corresponding activity), the remaining seed pixels related to this neuron have lowered PNR and local correlation. This helps avoid the duplicate initialization of the same neuron. Also, $P(\bm{x})$ and $L(\bm{x})$ have to be updated after each neuron is initialized, but since only a small area near the initialized neuron is affected, we can update these quantities locally to reduce the computational cost.  This procedure is repeated until the specified number of neurons have been initialized or no more candidate seed pixels exist. 

This initialization algorithm can greedily initialize the required number of neurons, but the subproblem of estimating $\hat{a}_i$ given $\hat{c}_i$ still has to deal with the large background activity in the residual matrix. We developed a simple method to remove this background and accurately initialize neuron shapes, described next.
We first crop a $(2l+1) \times (2l+1)$ square centered at $\bm{x}^*$ in the field of view (Figure \ref{fig:initialization}A-E). Then we compute the temporal correlation between the filtered traces of  pixel $x^*$ and all other pixels in the patch (Figure \ref{fig:initialization}D). We choose those pixels with small temporal correlations (e.g., 0.3) as the neighboring pixels that are outside of the neuron (the green contour in Figure \ref{fig:initialization}D). Next, we estimate the background fluctuations as the median values of these pixels for each frame in the raw data (Figure \ref{fig:initialization}E). We also select pixels that are within the neuron by selecting correlation coefficients larger than $0.7$, then $\hat{\bm{c}}_i$ is refined by computing the mean filtered traces of these pixels (Figure \ref{fig:initialization}E). Finally, we regress the raw fluorescence signal in each pixel onto three sources: the neuron signal (Figure \ref{fig:initialization}E), the local background fluctuation  (Figure \ref{fig:initialization}F), and a constant baseline. Our initial estimate of $\hat{a}_i$ is given by the regression weights onto $\hat{c}_i$ in Figure \ref{fig:initialization}F. 

\subsection{Interventions} 
We use iterative matrix updates to estimate model variables in CNMF-E. This strategy gives us the flexibility of integrating prior information on neuron morphology and temporal activity during the model fitting. The resulting interventions (which can in principle be performed either automatically or under manual control) can in turn lead to faster convergence and more accurate source extraction. We integrate $5$ interventions in our CNMF-E implementation. Following these interventions, we usually run one more iteration of matrix updates. 

\subsubsection{Merge existing components}
When a single neuron is split mistakenly into multiple components, a merge step is necessary to rejoin these components. If we can find all split components, we can superimpose all their spatiotemporal activities and run rank-$1$ NMF to obtain the spatial and temporal activity of the merged neuron. We automatically merge components for which the spatial and temporal components are correlated above certain thresholds. Our code also provides methods to manually specify neurons to be merged based on human judgement. 

\subsubsection{Split extracted components}
When highly correlated neurons are mistakenly merged into one component, we need to use spatial information to split into multiple components according to neurons' morphology.  Our current implementation of component splitting requires users to manually draw ROIs for splitting the spatial footprint of the extracted component. Automatic methods for ROI segmentation \citep{Apthorpe2016,Pachitariu2013} could be added as an alternative in future implementations.

\subsubsection{Remove false positives}
Some extracted components have spatial shapes that do not correspond to real neurons or temporal traces that do not correspond to neural activity. These components might explain some neural signals or background activity mistakenly. Our source extraction can benefit from the removal of these false positives. This can be done by manually examining all extracted components, or in principle automatically by training a classifier for detecting real neurons. The current implementation relies on visual inspection to exclude false positives. We also rank neurons based on their SNRs and set a cutoff to discard all extracted components that fail to meet this cutoff. As with the splitting step, removing false positives could also potentially use automated ROI detection algorithms in the future.  See \href{http://www.columbia.edu/~pz2230/videos/intervention_results.mp4}{S10 Video} for an example involving manual merge and delete operations.

\subsubsection{Pick undetected neurons from the residual}
If all neural signals and background are accurately estimated, the residual of the CNMF-E model $Y_{res} = Y -\hat{A}\hat{C}-\hat{B}$ should be relatively spatially and temporally uncorrelated. However, the initialization might miss some neurons due to large background fluctuations and/or high neuron density. After we estimate the background $\hat{B}$ and extract a majority of the neurons, those missed neurons have prominent fluorescent signals left in the residual. To select these undetected neurons from the residual $Y_{res}$, we use the same algorithm as for initializing neurons from the raw video data, but typically now the task is easier because the background has been removed. 

\subsubsection{Post-process the spatial footprints}
Each single neuron has localized spatial shapes and including this prior into the model fitting of CNMF-E, as suggested in \citep{Pnevmatikakis2016},  leads to better extraction of neurons spatial footprints. In the model fitting step, we constrain $A$ to be sparse and localized. These constraints do give us compact neuron shapes in most cases, but in some cases there are still some visually abnormal components detected. We include a heuristic automated post-processing step  after each iteration of updating spatial shapes (\ref{eq:opt_A}). For each extracted neuron $A(:, k)$, we first convert it to a 2D image and perform morphological opening to remove isolated pixels resulting from noise \citep{Haralick1987}. Next we label all connected components in the image and create a mask to select the largest component. All pixels outside of the mask in $A(:, i)$ are set to be $0$. This post-processing induces compact neuron shapes by removing extra pixels and helps avoid mistakenly explaining the fluorescence signals of the other neurons. 

\subsection{Pipeline, complexity analysis, and running time of CNMF-E} \label{sec:pipeline}

Our framework can be summarized in the following steps: 
\begin{enumerate}
  \item Initialize $\hat{A}, \hat{C}$ using the proposed initialization procedure.
  \item Solve problem (\ref{eq:opt_B}) for updates of $\hat{\bm{b}}_0$ and $\hat{B}^f$.
  \item Iteratively solve problem (\ref{eq:opt_A}) and (\ref{eq:opt_C}) to update $\hat{A}$ and $\hat{C}$.
  \item If desired, apply interventions to intermediate results.
  \item Repeat steps 2, 3, and  4 until the inferred components are stable.
\end{enumerate}
In practice, the estimation of the background $B$ (step 2) often does not vary greatly from  iteration to iteration and so this step usually can be run just once to save time.
In practice, we also use spatial and temporal decimation for improved speed, following \citep{Friedrich2016}. We first run the pipeline on decimated data to get good initializations, then we up-sample the results $\hat{A}, \hat{C}$ to the original resolution and run one iteration of steps (2-3) on the raw data. This strategy improves on processing the raw data directly because downsampling increases the signal to noise ratio and eliminates many false positives.

\subsubsection{Parameter selection}
Table \ref{table:pars} shows $5$ key parameters used in  CNMF-E. All of these parameters have interpretable meaning and can be easily picked within a broad range. The parameter $l$ controls the size of the spatial filter in the initialization step and is chosen as the diameter of a typical neuron in the FOV. As long as $l$ is much smaller than local background sources, the filtered data can be used for detecting seed pixels and then initializing neural traces. The distance between each seed pixel and its selected neighbors $l_n$ has to be larger than the neuron size $l$ and smaller than the spatial range of local background sources; in practice, this range is fairly broad. We usually set $l_n$ as $2l$.  To determine the thresholds $P_{\min}$ and $L_{\min}$, we first compute the correlation image and PNR image and then visually select very weak neurons from these two images. $P_{\min}$ and $L_{\min}$ are determined to ensure that CNMF-E is able to choose seed pixels from these weak neurons. Small $P_{\min}$ and $L_{\min}$ yield more false positive neurons, but they can be removed in the  intervention step. 
Finally, in practice, our results are not sensitive to the selection of the outlier parameter $\zeta$, thus we frequently set it as $10$.

\begin{table}[t!]
  \begin{tabular}{llll}
    \toprule
    \cmidrule{1-2}
    Name     & Description    
    & Default Values &  Used in\\
    \midrule
    $l$ & size of a typical neuron soma in the FOV
    & $30\mu m$ & Algorithm \ref{algo:init}\\
    $l_n$ & the distance between each pixel and its neighbors & $60\mu m$ &Problem (\ref{eq:opt_B})\\
    $P_{\min}$ & the minimum peak-to-noise ratio of seed pixels & $10$& Algorithm \ref{algo:init}\\
    $L_{\min}$ & the minimum local correlation of seed pixels & $0.8$& Algorithm \ref{algo:init}\\     
    $\zeta$ & the ratio between the outlier threshold and the noise & $10$ & Problem (\ref{eq:opt_B})\\
    \bottomrule
  \end{tabular} 
    \caption{Optional user-specified parameters.}
  \label{table:pars}
\end{table}

\subsubsection{Complexity analysis}
In step 1, the time cost is mainly determined by spatial filtering, resulting in $O(dl^2T)$ time. As for the initialization of a single neuron given a seed pixel, it is only ($O(l^2T)$). Considering the fact that the number of neurons is typically much smaller than the number of pixels in this data, the complexity for step 1 remains $O(dl^2T)$.  In step 2, the complexity of estimating $\hat{\bm{b}}_0$ is $O(dT)$ and estimating $\hat{B}^f$ scales linearly with the number of pixels $d$. For each pixel, the computational complexity for estimating $W_{i,:}$ is $O(l^2T)$. Thus the computational complexity in updating the background component is $O(dl^2T)$. In step 3, the computational complexities of solving problems (\ref{eq:opt_A}) and (\ref{eq:opt_C}) have been discussed in previous literature \citep{Pnevmatikakis2016} and they scale linearly with pixel number $d$ and time $T$, i.e., $O(dT)$. For the interventions, the one with the largest computational cost is picking undetected neurons from the residual, which is the same as the initialization step. Therefore, the computational cost for step 4 is $O(dl^2T)$. To summarize, the complexity for running CNMF-E is $O(dl^2T)$, i.e. the method scales linearly with both the number of pixels and the total recording time. 

\subsubsection{Running time}
To provide a sense of the running time of the algorithm, we timed the code on the simulation data shown in Figure \ref{fig:sim_compare}.
This dataset is $253\times 316$ pixels $\times 2000$ frames. PCA/ICA took $485$ seconds to  converge, using $250$ PCs and $220$ ICs. CNMF-E spent $67$ seconds for initialization, $48$ seconds for estimating the background, and $29$ seconds for updating spatial and temporal components, resulting in a total of $145$ seconds. Since the results already recovered the ground truth, we did not run more iterations.
The analyses were performed on a desktop with Intel Core i7-3770 CPU @ 3.40GHz and 12GB RAM running Ubuntu 14.04. Timing per iteration on real data examples was similar.
Our current implementation has not yet been highly optimized for speed. All algorithm steps can be easily parallelized; in the future we plan to pursue parallel approaches for speeding up the code.

\subsection{Simulation experiments}
\subsubsection{Details of the simulated experiment of Figure \ref{fig:bg}}
The field of view was $256\times 256$, with $1000$ frames. We simulated $50$ neurons whose shapes were simulated as spherical $2$-D Gaussian. The neuron centers were drawn uniformly from the whole FOV and the Gaussian widths $\sigma_x$ and $\sigma_y$ for each neuron was also randomly drawn from $\mathcal{N}\big(\frac{l}{4}, (\frac{1}{10}\frac{l}{4})^2\big)$, where $l=12$ pixels. Spikes were simulated from a Bernoulli process with probability of spiking per timebin $0.01$ and then convolved with a temporal kernel $g(t)=\exp(-t/\tau_d)-\exp(-t/\tau_r)$, with fall time $\tau_d= 6$ timebin and rise time $\tau_r = 1$ timebin. We simulated the spatial footprints of local backgrounds as $2$-D Gaussian as well, but the mean Gaussian width is $5$ times larger than the neurons' widths. As for the spatial footprint of the blood vessel in Figure \ref{fig:bg}A, we simulated a cubic function and then convolved it with a $2$-D Gaussian (Gaussian width=$3$ pixel). We use a random walk model to simulate the temporal fluctuations of local background and blood vessel. For the data used in Figure \ref{fig:bg}A-H, there were $23$ local background sources; for Figure \ref{fig:bg}I, we varied the number of background sources. 

We used the raw data to estimate the background in CNMF-E without subtracting the neural signals $\hat{A}\hat{C}$ in problem (\ref{eq:opt_B}).  We set $l_n=15$ pixels and left the remaining parameters at their default values. The plain NMF was performed using the built-in MATLAB function nnmf, which utilizes random initialization.  

\subsubsection{Details of the simulated experiment of Figure \ref{fig:fig_init} and Figure \ref{fig:sim_compare}}
We used the same simulation settings for both Figure \ref{fig:fig_init} and Figure \ref{fig:sim_compare}. The field of view was $253\times 316$ and the number of frames was $2000$. We simulated $200$ neurons using the same method as the simulation in Figure \ref{fig:bg}, but for the background we used the spatiotemporal activity of the background extracted using CNMF-E from real experimental data (data not shown). The noise level $\Sigma$ was also estimated from the data. When we varied the SNR in Figure\ref{fig:sim_compare}D-G, we multiplied $\Sigma$ with an SNR reduction factor. 

We set  
$l=12$ pixels to create the spatial filtering kernel. As for the thresholds used for determining seed pixels, we varied them for different SNR settings by visually checking the corresponding local correlation images and PNR images. The selected values were $L_{\min}=[0.9, 0.8, 0.8, 0.8 0.6, 0.6]$ and $P_{\min}=[30, 10, 10, 10, 8, 6]$ for different SNR reduction factors $[1,2,3,4,5,6]$. For PCA/ICA analysis, we set the number of PCs and ICs as $600$ and $300$ respectively. 

\subsection{\emph{In vivo} microendoscopic imaging and data analysis}

For all experimental data used in this work, we ran both CNMF-E and PCA/ICA. For CNMF-E, we chose parameters so that we initialized about 10-20\% extra components, which were then merged or deleted (some automatically, some under manual supervision) to obtain the final estimates.  Exact parameter settings are given for each dataset below.  For PCA/ICA, the number of ICs were selected to be slightly larger than our extracted components in CNMF-E (as we found this led to the best results for this algorithm), and the number of PCs was selected to capture over $90\%$ of the signal variance. The weight of temporal information in spatiotemporal ICA was set as $0.1$. After obtaining PCA/ICA filters, we again manually removed components that were clearly not neurons based on neuron morphology.  

We computed the SNR of extracted cellular traces to  quantitatively compare the performances of two approaches. For each cellular trace $\bm{y}$, we first computed its denoised trace $\bm{c}$ using the selected deconvolution algorithm (here, it is thresholded OASIS); then the SNR of $\bm{y}$ is 
\begin{equation}
	SNR = \frac{\|\bm{c}\|_2^2}{\|\bm{y}-\bm{c}\|_2^2}. 
\end{equation}
For PCA/ICA results, the calcium signal $\bm{y}$ of each IC is the output of its corresponding spatial filter, while for CNMF-E results, it is the trace before applying temporal deconvolution, i.e., $\hat{\bm{y}}_i$ in Eq. (\ref{eq:y_raw}). 



\subsubsection{Dorsal striatum data}
Expression of the genetically encoded calcium indicator GCaMP6f in neurons was achieved using a recombinant adeno-associated virus (AAV) encoding the GCaMP6f protein under transcriptional control of the synapsin promoter (AAV-Syn-GCaMP6f). This viral vector was packaged (Serotype 1) and stored in undiluted aliquots at a working concentration of $>1012$ genomic copies per ml at $-80^\circ$C until intracranial injection. $500\mu$l of AAV1-Syn-GCaMP6f was injected unilaterally into dorsal striatum ($0.6$ mm anterior to Bregma, $2.2$mm lateral to Bregma, $2.5$mm ventral to the surface of the brain). 1 week post injection, a $1$mm gradient index of refraction (GRIN) lens was implanted into dorsal striatum $\sim 300\mu$m above the center of the viral injection. 3 weeks after the implantation, the GRIN lens was reversibly coupled to a miniature 1-photon microscope with an integrated $475$nm LED (Inscopix).  Using nVistaHD Acquisition software, images were acquired at 30 frames per second with the LED transmitting $~0.1$ to $0.2$ mW of light while the mouse was freely moving in an open field arena. Images were down sampled to $10$Hz and processed into TIFFs using Mosaic software. All experimental manipulations were performed in accordance with protocols approved by the Harvard Standing Committee on Animal Care following guidelines described in the US NIH Guide for the Care and Use of Laboratory Animals. 

The parameters used in running CNMF-E were:  $l=15$ pixels, $l_n=30$ pixels, $\zeta=10$, $L_{\min}=0.7$, and $P_{\min}=7$. $513$ components were initialized from the raw data in the first pass before subtracting the background, and then additional components were initialized in a second pass. For this and the following experiments, the selected algorithm for updating spatial components was HALS \citep{Friedrich2016} and the method for deconvolving calcium traces was thresholded OASIS \citep{Friedrich2017}. Since the frame rate was relatively low ($10$ Hz), we used an AR($1$) model for the temporal traces. We used the same method selection in the following experimental datasets. We ran all $5$ types of interventions both automatically and manually. In the end, we obtained $550$ components. As for PCA/ICA analysis, the number of PCs and ICs were $2000$ and $700$ respectively.

\subsubsection{Prefrontal cortex data}
Cortical neurons were targeted by administering 2 microinjections of 300 ul of AAV-DJ-CamkIIa-GCaMP6s (titer: 5.3 x 1012, 1:6 dilution, UNC vector core) into the prefrontal cortex (PFC) (coordinates relative to bregma; injection 1: +1.5 mm AP, 0.6 mm ML, -2.4 ml DV; injection 2: +2.15 AP, 0.43 mm ML, -2.4 mm DV) of an adult male wild type (WT) mice. Immediately following the virus injection procedure, a 1 mm diameter GRIN lens implanted 300 um above the injection site (coordinates relative to bregma: +1.87 mm AP, 0.5 mm ML, -2.1 ml DV). After sufficient time had been allowed for the virus to express and the tissue to clear underneath the lens (~3 weeks), a baseplate was secured to the skull to interface the implanted GRIN lens with a miniature, integrated microscope (nVista, 473 nm excitation LED, Inscopix) and subsequently permit the visualization of Ca2+ signals from the PFC of a freely behaving mouse. The activity of PFC neurons were recorded at 15 Hz over a 10 min period (nVista HD Acquisition Software, Inscopix) while the test subject freely explored an empty novel chamber. Acquired data was spatially down sampled by a factor of 2, motion corrected, and temporally down sampled to 5 Hz (Mosaic Analysis Software, Inscopix). All procedures were approved by the University of North Carolina Institutional Animal Care and Use Committee (UNC IACUC).

The parameters used in running CNMF-E were: $l=12$ pixels, $l_n=24$ pixels, $\zeta=10$, $L_{\min}=0.65$, and $P_{\min}=10$. We first downsampled the data by $2$ both spatially and temporally. Then we applied CNMF-E to the downsampled data. There were $165$ components initialized in the first pass and we obtained $185$ components after running the whole CNMF-E pipeline. Then we up-sampled the results to the original solution and updated CNMF-E variables for one more iteration. For PCA/ICA, we used $275$ PCs and $250$ ICs.

\subsubsection{Ventral hippocampus data}
The calcium indicator GCaMP6f was expressed in ventral hippocampal-amygdala projecting neurons by injecting a retrograde canine adeno type 2-Cre virus (CAV2-Cre; from Larry Zweifel, University of Washington) into the basal amydala (coordinates relative to bregma:  -1.70 AP, 3.00mm ML, and -4.25mm DV from brain tissue at site), and a Cre-dependent GCaMP6f adeno associated virus (AAV1-flex-Synapsin-GCaMP6f, UPenn vector core) into ventral CA1 of the hippocampus (coordinates relative to bregma: -3.16mm AP, 3.50mm ML, and -3.50mm DV from brain tissue at site). A 0.5mm diameter GRIN lens was then implanted over the vCA1 subregion and imaging began 3 weeks after surgery to allow for sufficient viral expression. Mice were then imaged with Inscopix miniaturized microscopes and nVistaHD Acquisition software as described above; images were acquired at 15 frames per second while mice explored an anxiogenic Elevated Plus Maze arena. Videos were motion corrected and spatially downsampled using Mosaic software. All procedures were performed in accordance with protocols approved by the New York State Psychiatric Institutional Animal Care and Use Committee following guidelines described in the US NIH Guide for the Care and Use of Laboratory Animals. 

The parameters used in running CNMF-E were:  $l=15$ pixels, $l_n=30$ pixels, $\zeta=10$, $L_{\min}=0.85$, and $P_{\min}=15$. We first temporally downsampled the data by $2$. Then we applied CNMF-E to the downsampled data. There were $59$ components initialized. We merged most of these components and deleted false positives. In the end, there were $22$ components left. The intermediate results before and after each manual intervention are shown in \href{http://www.columbia.edu/~pz2230/videos/intervention_results.mp4}{S10 Video}. Then we up-sampled the results to the original solution and updated CNMF-E for one more iteration (steps 2$\&$3). As for PCA/ICA analysis, the number of PCs and ICs are $100$ and $40$ respectively. 

\subsubsection{BNST data with footshock}
Calcium indicator GCaMP6s was expressed within CaMKII-expressing neurons in the BNST by injecting the recombinant adeno-associated virus AAVdj-CaMKII-GCaMP6s (packaged at UNC Vector Core) into the anterior dorsal portion of BNST (coordinates relative to bregma: 0.10mm AP, -0.95mm ML, -4.30mm DV). A 0.6 mm diameter GRIN lens was implanted above the injection site within the BNST. As described above, images were acquired using a detachable miniature 1-photon microscope and nVistaHD Acquisition Software (Inscopix).  Images were acquired at 20 frames per second while the animal was freely moving inside a sound-attenuated chamber equipped with a house light and a white noise generator (Med Associates).  Unpredictable foot shocks were delivered through metal bars in the floor as an aversive stimulus during a 10-min session.  Each unpredictable foot shock was 0.75 mA in intensity and 500 ms in duration on a variable interval (VI-60). As described above, images were motion corrected, downsampled and processed into TIFFs using Mosaic Software. These procedures were conducted in adult C57BL/6J mice (Jackson Laboratories) and in accordance with the Guide for the Care and Use of Laboratory Animals, as adopted by the NIH, and with approval from the Institutional Animal Care and Use Committee of the University of North Carolina at Chapel Hill (UNC).  

The parameters used in running CNMF-E were:  $l=15$ pixels, $l_n=30$ pixels, $\zeta=10$, $L_{\min}=0.9$, and $P_{\min}=15$. There were $150$ components initialized and there were $126$ components left after running the whole pipeline.  The number of PCs and ICs in PCA/ICA were $200$ and $150$, respectively. 

\subsection{Code availability}
All analysis was performed with custom-written MATLAB code. MATLAB implementations of CNMF-E algorithm can be freely downloaded on \url{https://github.com/zhoupc/CNMF_E}.

\section{Supporting information}
{\bf S1 Video. An example of typical microendoscopic data.}  The video was recorded in dorsal striatum; experimental details can be found above.\\
\href{http://www.columbia.edu/~pz2230/videos/example_microendoscopic_data.mp4}{\bf MP4}

\noindent {\bf S2 Video. \bf Comparison of CNMF-E with rank-1 NMF in estimating background fluctuation in simulated data.} Top left: the simulated fluorescence data in Figure \ref{fig:bg}. Bottom left: the ground truth of neuron signals in the simulation. Top middle: the estimated background from the raw video data (top left) using CNMF-E. Bottom middle: the residual of the raw video after subtracting the background estimated with CNMF-E. Top right and top bottom: same as top middle and bottom middle, but the background is estimated with rank-1 NMF. \\
\href{http://www.columbia.edu/~pz2230/videos/background_comparison.mp4}{\bf MP4}

\noindent {\bf S3 Video. \bf Initialization procedure for the simulated data in Figure \ref{fig:fig_init}.} Top left: correlation image of the filtered data. Red dots are centers of initialized neurons. Top middle: candidate seed pixels (small red dots) for initializing neurons on top of PNR image. The large red dot indicates the current seed pixel. Top right: the correlation image surrounding the selected seed pixel or the spatial footprint of the initialized neuron. Bottom: the filtered fluorescence trace at the seed pixel or the initialized temporal activity (both raw and denoised). \\
\href{http://www.columbia.edu/~pz2230/videos/sim_initialization.mp4}{\bf MP4}

\noindent {\bf S4 Video. \bf The results of CNMF-E in demixing simulated data in Figure \ref{fig:sim_compare} (SNR reduction factor=1).} Top left: the simulated fluorescence data. Bottom left:  the estimated background. Top middle: the residual of the raw video (top left) after subtracting the estimated background (bottom left).  Bottom middle: the denoised neural signals. Top right: the residual of the raw video data (top right) after subtracting the estimated background (bottom left) and denoised neural signal (bottom middle). Bottom right: the ground truth of neural signals in simulation. \\
\href{http://www.columbia.edu/~pz2230/videos/sim_snr1_demixing.mp4}{\bf MP4}

\noindent {\bf S5 Video. \bf The results of CNMF-E in demixing the simulated data in Figure \ref{fig:sim_compare} (SNR reduction factor=6).} Conventions as in previous video. \\
 \href{http://www.columbia.edu/~pz2230/videos/sim_snr1_demixing.mp4}{\bf MP4}

\noindent {\bf S6 Video. \bf The results of CNMF-E in demixing dorsal striatum data.} Top left: the recorded fluorescence data. Bottom left:  the estimated background. Top middle: the residual of the raw video (top left) after subtracting the estimated background (bottom left).  Bottom middle: the denoised neural signals. Top right: the residual of the raw video data (top right) after subtracting the estimated background (bottom left) and denoised neural signal (bottom middle). Bottom right: the denoised neural signals while all neurons' activity are coded with pseudocolors. \\
\href{http://www.columbia.edu/~pz2230/videos/striatum_demixing.mp4}{\bf MP4}

\noindent {\bf S7 Video. \bf The results of CNMF-E in demixing PFC data.} Conventions as in previous video. \\
\href{http://www.columbia.edu/~pz2230/videos/pfc_demixing.mp4}{\bf MP4}

\noindent {\bf S8 Video. \bf Comparison of CNMF-E with PCA/ICA in demixing overlapped neurons in Figure \ref{fig:pfc}G.} Top left: the recorded fluorescence data. Bottom left: the residual of the raw video (top left) after subtracting the estimated background using CNMF-E. Top middle and top right: the spatiotemporal activity and temporal traces of three neurons extracted using CNMF-E. Bottom middle and bottom right:  the spatiotemporal activity and temporal traces of three neurons extracted using PCA/ICA. \\
 \href{http://www.columbia.edu/~pz2230/videos/pfc_overlapping.mp4}{\bf MP4}

\noindent {\bf S9 Video. \bf The results of CNMF-E in demixing ventral hippocampus data.} Conventions as in S6 Video. \\
\href{http://www.columbia.edu/~pz2230/videos/sparse_demixing.mp4}{\bf MP4}

\noindent {\bf S10 Video. \bf Extracted spatial and temporal components of CNMF-E at different stages (ventral hippocampal dataset).} After initializing components, we ran matrix updates and interventions in automatic mode, resulting in $32$ components in total. In the next iteration, we manually deleted $6$ components and automatically merged neurons as well. In the last iterations, $4$ neurons were merged into $2$ neurons with manual verifications. The correlation image in the top left panel is computed from the background-subtracted data in the final step.\\
\href{http://www.columbia.edu/~pz2230/videos/intervention_results.mp4}{\bf MP4}

\noindent {\bf S11 Video. \bf The results of CNMF-E in demixing BNST data.} Conventions as in S6 Video. \\
\href{http://www.columbia.edu/~pz2230/videos/footshock_demixing.mp4}{\bf MP4}

\section{Acknowledgments}
We would like to thank CNMF-E users who received early access to our package and provided tremendously helpful feedback and suggestions, especially James Hyde, Jesse Wood, and Sean Piantadosi in Susanne Ahmari's lab in University of Pittsburgh, Andreas Klaus in Rui Costa's Lab in the Champalimaud Neurobiology of Action Laboratory, Suoqin Jin in Xiangmin Xu's lab at University of California - Irvine, Conor Heins at the National Institute of Drug Abuse, Chris Donahue in Anatol Kreitzer's lab at University of California - San Francisco, Xian Zhang in Bo Li's lab at Cold Spring Harbor Laboratory, Emily Mackevicius in Michale Fee's lab at Massachusetts Institute of Technology, Courtney Cameron and Malavika Murugan in Ilana Witten's lab at Princeton University, Pranav Mamidanna in Jonathan Whitlock's lab at Norwegian University of Science and Technology, and Milekovic Tomislav in Gregoire Courtine's group at EPFL. We thank Eftychios Pnevmatikakis and Johannes Friedrich for helpful discussions on solving CNMF problems.  We also thank Andreas Klaus  and Johannes Friedrich for valuable comments on the manuscript. 

\section{Additional information}
\clearpage
\subsection{Funding}
\begin{table}[!h]
\begin{center}
\begin{tabular}{L{0.3\textwidth} L{0.3\textwidth} L{0.33\textwidth}}
\hline
Funder & Grant reference& Author\\ [0.5ex] 
\hline
 National Institute of Mental Health (NIMH) 
 & 2R01MH064537 &  Robert E. Kass, Pengcheng Zhou\\ 
\cline{2-3}
& R37 MH068542 & Rene Hen, Jessica C. Jimenez\\
 \cline{2-3}
& R01 MH083862 &  Rene Hen, Jessica C. Jimenez\\ 
 \cline{2-3}
&  R01 MH108623 & Mazen A. Kheirbek\\
\hline
 National Institute on Drug Abuse (NIDA) & R90 DA023426 & Pengcheng Zhou\\ 
\cline{2-3}
& R01 DA038168 & Garret D Stuber, Jose Rodriguez-Romaguera  \\
 \hline 
Intelligence Advanced Research Projects Activity (IARPA) 
 & DoI/IBC D16PC00007 & Pengcheng Zhou\\ 
\cline{2-3}
&  DoI/IBC D16PC00003 & Liam Paninski\\ 
\hline
Defense Advanced Research Projects Agency (DARPA) & N66001-15-C-4032 & Liam Paninski\\
 \hline
 Army Research Office (ARO)  & MURI W911NF-12-1-0594 & Liam Paninski \\
\hline 
National Institute of Biomedical Imaging and Bioengineering (NIBIB) & R01 EB22913 & Liam Paninski \\ 
\hline
Eunice Kennedy Shriver National Institute of Child Health and Human Development (NICHD) & T32 HD040127 & Shanna L. Resendez \\
 \cline{2-3}
& U54-HD079124 & Garret D Stuber \\
 \hline 
 Howard Hughes Medical Institute (HHMI) &  Gilliam Fellowship for Advanced Study & Jessica C. Jimenez\\ 
 \hline  
 National Institute on Aging (NIA) & R01 AG043688 &  Rene Hen, Jessica C. Jimenez\\
 \hline  
 New York State Stem Cell Science (NYSTEM)&NYSTEM-C029157 & Rene Hen, Jessica C. Jimenez \\
 \hline  
 Hope For Depression Research Foundation (HDRF) & RGA-13-003 &Rene Hen,  Jessica C. Jimenez \\
 \hline  
 Canadian Institutes of Health Research (CIHR) & Doctoral Foreign Study Award &  Shay Q. Neufeld\\
\hline 
 The Simons Foundation & & Garret D Stuber \\
  \hline 
 International Mental Health Research Organization (IMHRO) &  Rising Star Award & Mazen A. Kheirbek\\
  \hline 
National Institute of Neurological Disorders and Stroke (NINDS) & U01NS094191 &  Bernardo L. Sabatini\\ 
\hline
 \multicolumn{3}{L{1\textwidth}}{The funders had no role in study design, data collection and interpretation, or the decision to submit the work for publication.} \\
 \hline
 \end{tabular}
 \end{center}
\end{table}

\bibliographystyle{agsm}
\bibliography{references}
\end{document}